\documentclass[fleqn,usenatbib]{mnras}

\usepackage{newtxtext,newtxmath}
\usepackage[T1]{fontenc}
\usepackage{ae,aecompl}

\usepackage{graphicx}	% Including figure files
\usepackage{amsmath}	% Advanced maths commands
\usepackage{amssymb}	% Extra maths symbols
\usepackage{multirow}
\usepackage{booktabs}

\newcommand{\kms}{km~s$^{-1}$}
\newcommand{\cmN}{cm$^{-2}$}

\newcommand{\lam}{$\lambda$}

\newcommand{\ha}{H$\alpha$}
\newcommand{\hb}{H$\beta$}
\newcommand{\hg}{H$\gamma$}
\newcommand{\hd}{H$\delta$}
\newcommand{\lya}{\mbox{Ly$\alpha$}}

\newcommand{\hi}{\mbox{H\,{\sc i}}}
\newcommand{\hii}{\mbox{H\,{\sc ii}}}
\newcommand{\hei}{\mbox{He\,{\sc i}}}
\newcommand{\heii}{\mbox{He\,{\sc ii}}}

\newcommand{\civ}{\mbox{C\,{\sc iv}}}
\newcommand{\ciii}{\mbox{C\,{\sc iii}}}
\newcommand{\cii}{\mbox{C\,{\sc ii}}}
\newcommand{\ci}{\mbox{C\,{\sc i}}}
\newcommand{\siiv}{\mbox{Si\,{\sc iv}}}

\newcommand{\siii}{\mbox{Si\,{\sc ii}}}
\newcommand{\siv}{\mbox{S\,{\sc iv}}}

\newcommand{\sii}{\mbox{S\,{\sc ii}}}
\newcommand{\nv}{\mbox{N\,{\sc v}}}

\newcommand{\niii}{\mbox{N\,{\sc iii}}}
\newcommand{\nii}{\mbox{N\,{\sc ii}}}
\newcommand{\nti}{\mbox{N\,{\sc i}}}

\newcommand{\ovi}{\mbox{O\,{\sc vi}}}

\newcommand{\oiii}{\mbox{O\,{\sc iii}}}

\newcommand{\oi}{\mbox{O\,{\sc i}}}

\newcommand{\feii}{\mbox{Fe\,{\sc ii}}}
\newcommand{\feiii}{\mbox{Fe\,{\sc iii}}}
\newcommand{\mgi}{\mbox{Mg\,{\sc i}}}
\newcommand{\mgii}{\mbox{Mg\,{\sc ii}}}
\newcommand{\caii}{\mbox{Ca\,{\sc ii}}}
\newcommand{\aliii}{\mbox{Al\,{\sc iii}}}

\newcommand{\pv}{\mbox{P\,{\sc v}}}

\newcommand{\nai}{\mbox{Na\,{\sc i}}}
\newcommand{\naii}{\mbox{Na\,{\sc ii}}}
\newcommand{\nkii}{\mbox{Ni\,{\sc ii}}}
\newcommand{\cloudy}{\textsc{cloudy}}

\title[A High-Column Density Outflow from PG~1411+442]{On the Emergence of Thousands of Absorption Lines in the Quasar PG~1411+442: A Clumpy High-Column Density Outflow from the Broad Emission-Line Region?}
\author[Fred Hamann et al.]{
Fred Hamann$^{1}$\thanks{e-mail: fhamann@ucr.edu},
Todd M. Tripp$^{2}$,
David Rupke$^{3}$,
Sylvain Veilleux$^{4,5,6}$
\\
$^{1}$Department of Physics \& Astronomy, University of California, Riverside, CA 92507, USA \\
$^{2}$Department of Astronomy, University of Massachussetts, Amherst, MA 01003, USA\\
$^3$Department of Physics, Rhodes College, Memphis, TN 38112, USA\\
$^4$Department of Astronomy, University of Maryland, College Park, MD 20742, USA\\
$^5$Joint Space-Science Institute, University of Maryland, College Park, MD 20742, USA\\
$^6$Institute of Astronomy and Kavli Institute for Cosmology Cambridge, University of Cambridge, Cambridge, CB3 0HA, United Kingdom}

\date{Accepted XXX. Received YYY; in original form ZZZ}

\pubyear{2015}

\begin{document}
\label{firstpage}
\pagerange{\pageref{firstpage}--\pageref{lastpage}}
\maketitle

\begin{abstract}
Quasar outflows are fundamental components of quasar environments that might play an important role in feedback to galaxy evolution. We report on the emergence of a remarkable new outflow absorption-line system in the quasar PG1411+442 (redshift $\sim$0.089) detected in the UV and visible with the Hubble Space Telescope Cosmic Origins Spectrograph and the Gemini Multi-Object Spectrograph, respectively. This new ``transient'' system contains thousands of lines, including \feii\ and \feii* from excited states up to 3.89 eV, \hi* Balmer lines, \nai~D \lam 5890,5896, and the first detection of \hei* \lam 5876 in a quasar. The transient absorber is spatially inhomogeneous and compact, with sizes $\lesssim$0.003~pc, based on covering fractions on the quasar continuum source ranging from $\sim$0.45 in strong UV lines to $\sim$0.04 in \nai~D. \cloudy\ photoionization simulations show that large total column densities $\log N_{\rm H}({\rm cm}^{-2})\gtrsim23.4$ and an intense radiation field $\lesssim$0.4~pc from the quasar are needed to produce the observed lines in thick zones of both fully-ionised and partially-ionised gas. The densities are conservatively $\log n_{\rm H}({\rm cm}^{-3})\gtrsim7$ based on \feii*, \hi*, and \hei* but they might reach $\log n_{\rm H}({\rm cm}^{-3})\gtrsim10$ based on \nai~D. The transient lines appear at roughly the same velocity shift, v$\;\sim-1900$~\kms, as a ``mini-BAL'' outflow detected previously, but with narrower Doppler widths, $b\sim100$~\kms, and larger column densities in more compact outflow structures. We propose that the transient lines identify a clumpy outflow from the broad emission-line region that, at its current speed and location, is still gravitationally bound to the central black hole.
\end{abstract}


%%%%%%%%%%%%%%%%%%%%%%%%%%%%%%%%%%%%%%%%%%%%%%%%%%

%%%%%%%%%%%%%%%%% BODY OF PAPER %%%%%%%%%%%%%%%%%%

\section{Introduction}

Quasar accretion-disk outflows might play an important role in galaxy evolution by disrupting star formation and ejecting interstellar gas and dust from the host galaxies \citep[e.g.,][]{diMatteo05, DeBuhr12, Faucher12, Cicone14, Hopkins10, Hopkins16, Veilleux13b, Rupke13, Rupke17}. The main direct tracers of these outflows are blueshifted broad absorption lines (BALs) in quasar spectra. However, more than 50 years after the discovery of BALs \citep{Lynds67}, we still have a poor understanding of basic outflow properties such as their locations, spatial structure, densities, column densities, kinetic energies available for ``feedback'' to the host galaxies, and the physical processes that control the launching and evolution of the outflows.

Densities and column densities are fundamental parameters that affect our estimates of the outflow locations and kinetic energies. The common picture of BAL outflows is that they are launched from quasar accretion disks and the lines we measure form within a few pc of the central black holes \citep[e.g.,][]{Murray95, Murray97, Proga04, Everett05, Kazanas12, Matthews16}. Some observational studies report small distances consistent with this picture \citep[e.g.,][and refs. therein]{Wampler95, deKool02c, Hall11, Capellupo13, Zhang15b, Muzahid16, Shi16b, Moravec17, Hamann18}, while others derive much larger radial distances up to several kpc \citep{deKool01, Moe09, Dunn10, Chamberlain15, Chamberlain15b, Arav18}. The wide range in reported radial distances, $R$, is important for feedback considerations because the derived outflow masses and kinetic energies scale like $R^2$, while the kinetic energy rates (divided by an average flow time $R$/v) scale like $R$ \citep[e.g.,][]{Moravec17}. 

Large BAL outflow distances usually follow from low densities and photoionisation constraints combined with low-to-moderate levels of ionisation in the outflows \citep[see also][for similar results for narrow ``associated'' absorption lines]{Sargent82, Tripp96, Srianand00, Hamann01, Finn14, Chen18}. 
However, the actual distances could be smaller if the outflows have large column densities to provide radiative shielding for lower ions within the outflows. Unfortunately, column densities are notoriously difficult to measure from BAL spectra because partial covering of the emission source leads to unabsorbed flux filling in the absorption-line troughs. This makes the observed line strengths unreliable indicators of the true optical depths and column densities. A common strategy to diagnose partial covering and constrain the true column densities is to examine line ratios in doublets or multiplets with known optical depth ratios, such as \civ\ \lam 1548,1551 \citep{Hamann97, Hamann01, Hamann11, Ganguly99, Gabel05, Arav08}. Another strategy is to examine lines arising from low-abundance ions like \pv\ \lam 1118,1128 or highly-excited states like \hei* \lam 3889, 10830 that require large column densities to produce significant absorption \citep{Hamann18, Leighly11, Borguet12, Ji15, Liu15, Shi16, Wildy16, Capellupo17}. 

We are involved in a program to address these questions using new rest-frame UV spectra of six low-redshift quasars ($0.07 \lesssim z_{e} \lesssim 0.36$) obtained with the Cosmic Origins Spectrograph (COS) on the Hubble Space Telescope (HST). The quasars were known already to have broad outflow absorption lines in previous HST-COS spectra \citep[][Veilleux et al., in prep.]{Danforth16}. These quasars are valuable for outflow studies because 1) they fill a largely-unexplored niche between luminous quasars with strong BALs reaching speeds of 0.1$c$ or 0.2$c$ and low-luminosity Seyfert 1 galaxies with exclusively narrower outflow lines at lower speeds, 2) the low redshifts minimize contamination in the \lya\ forest for lines at wavelengths $\lesssim$1216 \AA , and 3) the outflow lines are relatively narrow to minimize blending problems. In particular, the lines are ``mini-BALs'' instead of BALs because their velocity widths are below the minimum of $\gtrsim$2000 \kms\ generally used for BALs \citep{Weymann91}. We will present the full results of this study in a follow-up paper by Hamann et al. (in prep., hereafter HTRV2). 

Here we report on the remarkable emergence of a rich system of several thousand narrow absorption lines in one of these quasars, PG~1411+442 at $z_e\approx 0.08982$. This exotic new outflow system, hereafter the ``transient'' system, appeared across the UV through visible spectrum at roughly the same velocity shift, v$\,\sim -1900$ \kms , as the mini-BALs measured previously. We argue below that the transient lines identify another component in the overall clumpy inhomogeneous outflow from PG~1411+442, e.g., with higher densities and larger column densities in a spatially smaller structure than the mini-BAL gas. 

The transient outflow system includes exotic features like excited-state \hei* \lam 5876 absorption and several high-energy \feii* multiplets not reported previously in an AGN outflow. But its rich spectrum resembles low-ionisation outflows observed in other AGN such as the Seyfert 1 galaxy NGC~4151 \citep{Crenshaw00, Kraemer01} and a subset of BAL quasars called ``FeLoBALs'' defined by \feii\ and \feii* absorption in addition to common higher-ionisation lines like \civ\ \citep{Hazard87, Wampler95, Hall02, Hall03, deKool02b, Trump06}. The unique variety of well-measured lines in PG~1411+442 provides a unique opportunity to place firm constraints on the outflow physical properties, with potentially broad implications for the nature of the quasar outflows launched from the accretion disk. 

After our initial HST-COS observations, we continued monitoring PG~1411+442 in both the UV and visible to track the evolution of the transient outflow lines. Those monitoring results will also be reported in HTRV2. Here we discuss the emergence of the transient lines with a detailed analysis of the absorber physical conditions in 2015 when the lines were strongest in our data. Section 2 describes the relevant observations and data reductions. Section 3 summarizes basic properties of the quasar that are useful for our analysis. Section 4 presents the spectra and provides an overview of the outflow absorption-line features. Section 5 describes the line identifications,  line measurements, and estimated the physical properties derived from the line strengths, line ratios, and comparisons to \cloudy\ photoionisation simulations. Section 6 provides a brief summary and discussion of the results.

Throughout this paper, we adopt a cosmology with $H_o = 69.6$ \kms\ Mpc$^{-1}$, $\Omega_M = 0.286$ and $\Omega_{\lambda}=0.714$ \citep{Bennett14}, and we use the online cosmology calculator developed by \cite{Wright06} for the luminosity distance. 

\section{Observations \& Data Reductions}

Table 1 summarizes the observations of PG~1411+442 discussed in this paper. This includes UV spectra obtained with HST-COS \citep{Green12} in 2011 and 2015, and a ground-based visible spectrum obtained with the Gemini Multi-Object Spectrograph \citep[GMOS,][]{Allington-Smith02, Hook04} in 2015 approximately 0.3 years (in the quasar frame) after the 2015 HST-COS observation. 

\begin{table}
	\centering
	\caption{Log of Observations. The columns give the observation dates, telescope and instrument used, spectrograph grating, observer-frame wavelength ranges, and approximate resolutions $R \equiv \lambda /\Delta\lambda$ near the center of the spectral coverage.
}
	\begin{tabular}{lcccc} % four columns, alignment for each
		\hline
		Date & Tel-Instr& Grating& $\lambda$ range& $R$\\
		& & & (\AA )& \\
		\hline
	\vspace{4pt}
		2011 Oct 23& HST-COS & G130M& 1065-1365& $\sim$12,000  \\
		2015 Feb 12& HST-COS& G130M& 930-1236& $\sim$10,000 \\
		& & G130M& 1150-1450& $\sim$18,000\\
%	\vspace{4pt}
		& & G160M& 1405-1775& $\sim$18,000\\
%		2016 Apr 16& HST-COS & G130M& 930-1236& $\sim$10,000\\
%		& & G130M& 1150-1450& $\sim$18,000\\
%		& & G160M& 1405-1775& $\sim$18,000\\
	\multicolumn{5}{c}{\dotfill}\\
		\multirow{2}{*}{2015 June 22,23}& Gemini& \multirow{2}{*}{B600}& \multirow{2}{*}{4615-7472}& \multirow{2}{*}{$\sim$5000}\\
		& -GMOS& & & \\
%	\vspace{3pt}
%		\multirow{2}{*}{2016 Feb 05}& Gemini& \multirow{2}{*}{---}& \multirow{2}{*}{4100-9800}& \multirow{2}{*}{$\sim$40,000}\\
%		& /GRACES\\
		\hline
	\end{tabular}
\end{table}

\subsection{UV spectra: HST-COS}

The UV spectra obtained with HST-COS were reduced using the standard CALCOS pipeline (version 3.0.0) with default parameters to perform basic procedures including the geometric distortion correction, flat-fielding to remove the features imprinted on the spectra by the grid wires in the COS detector, wavelength calibration, and extraction of the one-dimensional spectra (see the COS Data Handbook, 2018, for more details). After extracting the one-dimensional COS spectra from each exposure, we refined the wavelength alignment of the exposures by applying an offset derived from comparison of interstellar absorption lines that are well-detected in the 1d spectra, and then we tried two methods for coadding to the data to produce the final spectrum.  First, since COS is a photon-counting instrument with very low backgrounds and the data were recorded time-tag mode, we coadded the data using the software described by \cite{Meiring11}, which correctly tracks the number of source and background counts in each pixel and thereby provides a robust estimate of the flux uncertainties based on counting statistics.  However, this software is not designed to easily apply the absolute flux calibration, so we also tried coadding the flux-calibrated pipeline spectra with the weighting scheme of \cite{Tripp01}. In the end, we found that the two methods produced nearly identical spectra and flux uncertainties, so we opted to use the latter method because the absolute flux is of interest in this study.

\subsection{Visible spectra: Gemini-GMOS IFU}

The optical spectra were acquired with the integral field unit (IFU) of Gemini-GMOS in one-slit mode as part of program GN-2015A-DD-9. The exposure was two hours, and the delivered image quality was 0\farcs8. We used the Gemini IRAF software package (version 1.12) to reduce the data, supplemented by revision 114 of the IFUDR GMOS, IFSRED \citep{Rupke14a}, and QFitsView \citep{Ott12}. The basic steps are outlined in \cite{Rupke13} and \cite{Rupke15}. Instrumental scattered light was also removed from the flat fields and the data prior to extraction. However, the scattered light could not be completely removed because the dithering put the quasar near the edge of the FOV along the short axis of the IFU. This concentrated the quasar light near the edge of fiber blocks in the raw data, which meant that consecutive interblock regions sampled the scattered light unevenly. The result is a stripe of residual scattered light through the center of the data cube.

The quasar nuclear spectrum was extracted from a circular aperture of diameter 1\farcs0, slightly larger than the seeing disk. It was then modeled with a quasar and host galaxy template using IFSFIT \citep{Rupke14b}. The quasar template was the spectrum within the central 0.1$\times$0.1 arcsecond spaxel of the data cube, and the starlight template was the integrated starlight from the host, derived from the iterative spectral decomposition method of \cite{Rupke17}. Extrapolating the starlight measured in the host galaxy across the central spaxels dominated by quasar light indicates that the starlight contribution to the nuclear 1.0 arcsecond ground-based aperture is negligible, $<$1 per cent.

Negligible starlight contributions are also expected from an HST image of PG~1411+442 obtained in visible light by \cite{Bentz09}. This image, with $\sim$0.1 arcsecond resolution, shows that the central quasar is $\sim$3.7 times brighter than the \textit{integrated} flux from the entire host galaxy at 5100 \AA\ (rest). We also find no evidence for stellar absorption lines in the nuclear Gemini-GMOS spectrum. We compare this PG~1411+442 spectrum to the composite spectrum of SDSS quasars in \cite[][measured through 3 arcsecond fibers]{VandenBerk01}, which does contain stellar absorption lines such as \nai~D \lam 5890,5896 that those authors use to estimate typical starlight contributions of 7 to 15 percent at $\sim$5000 \AA . Our upper limit on the {\it stellar} \nai~D lines in the Gemini-GMOS spectrum of PG~1411+442 is at least 4 to 5 times weaker than the lines measured by \cite{VandenBerk01}, indicating, again, that starlight does not contribute significantly to our visible spectrum of PG~1411+442. 

Starlight contributions to the UV spectra should be even less because 1) the stars become relatively fainter than the quasar at short wavelengths, and 2) the HST-COS extraction aperture was much smaller Gemini-GMOS to include fewer stars. This is confirmed by inspection of the UV acquisition images obtained by HST-COS, which indicate that the host galaxy is negligibly faint compared to the quasar within our UV extraction aperture. We also note that the \civ\ \lam 1548,1551 mini-BAL in the HST-COS spectrum extends down to a normalized flux of $\sim$7 percent, which places a firm upper limit on the starlight contribution at those wavelengths (see HTRV2).

Other results from the GMOS-IFU observations are reported in \cite{Rupke17}. They find a systemic redshift based on the stellar velocity field of  $z\approx 0.0898$, consistent with the redshift we derive from the [\oiii] \lam 5007 emission line $z_e = 0.08982$ (Section 3). The host galaxy shows extended \nai~D absorption at velocities of 120 and $-$1300 km/s with respect to this velocity. The redshifted absorption extends to the edges of the GMOS field of view (at least 6 kpc). The blue shifted absorption extends roughly 3 kpc from the nucleus/quasar in several directions, indicating a high-velocity neutral galactic wind \citep[see][for more discussion]{Rupke17}.

\section{Quasar Properties}

Here we derive basic properties of PG~1411+442 useful for our analysis below. We measured the systemic redshift from the [\oiii ] \lam 5007 (5006.843 \AA ) emission line in our 2015 Gemini-GMOS spectrum after shifting to the heliocentric frame. The profile of the [\oiii] line has a significant asymmetry at its base, with a blue excess that is probably related to outflows in an extended low-density environment. To avoid this blueshifted component, we measure the centroid wavelength of only the upper 50 percent of the line profile. The redshift thus derived is $z_e = 0.08982$. The formal uncertainty due to noise in the spectrum is negligibly small ($<$1 \kms ) compared to potential offsets in the [OIII] line due to gas flows \citep[still typically $\lesssim$50 \kms ,][and refs. therein]{Shen16, Shen16b}. 

We estimate the luminosity of PG~1411+442 at rest-frame wavelength 1500 \AA\ from an approximate average of the observed fluxes in 2011, 2015 and 2016 (the latter is discussed in HTRV2). The observed UV flux was lowest in 2015, higher by a factor of $\sim$2 in 2016, and roughly midway between those measurements in 2011. Thus we adopt the measured value 
%\footnote{The rest wavelength 1500 \AA\ was just outside the COS spectral coverage in 2011. We estimate the flux at that wavelength by extrapolation, guided by the measurements in 2015 and 2016 that bracket the 2011 fluxes.} 
in 2011: $F_{\lambda} (1500) \approx 1.5 \times 10^{-14}$ ergs s$^{-1}$ cm$^{-2}$ \AA$^{-1}$. This flux corresponds to $\lambda L_{\lambda}(1500) \sim 5 \times 10^{44}$ ergs s$^{-1}$ at the quasar redshift, and a bolometric luminosity of $L \sim 2 \times 10^{45}$ ergs s$^{-1}$ based on a bolometric correction factor of $L \sim 4.0\, \lambda L_{\lambda}$(1500) that is consistent with previous continuum/photometric observations of PG~1411+442 (Section 5.3). This luminosity is very similar to the value $L\sim 2.3 \times 10^{45}$ ergs s$^{-1}$ derived by \cite{Netzer07a} from visible and infrared measurements. 

PG~1411+442 has been targeted in \hb\ reverberation studies to estimate its black hole mass and the radial size of the broad emission line region \citep[BLR, e.g.,][]{Peterson04, Bentz09}. The most recent studies indicate $M_{BH} \approx (3.5\pm 1.2)\times 10^8$ M$_{\odot}$ and $R_{BLR} \approx 0.10\pm 0.05$ pc, respectively \citep[from][and the online database maintained by Misty Bentz, http://www.astro.gsu.edu/AGNmass/]{Bentz15}.  The mass estimate combined with the luminosity above indicates a relatively small Eddington ratio of $L/L_E \sim 0.04$. 

\section{Outflow Absorption Line Overview}

\begin{figure}
\includegraphics[width=\columnwidth]{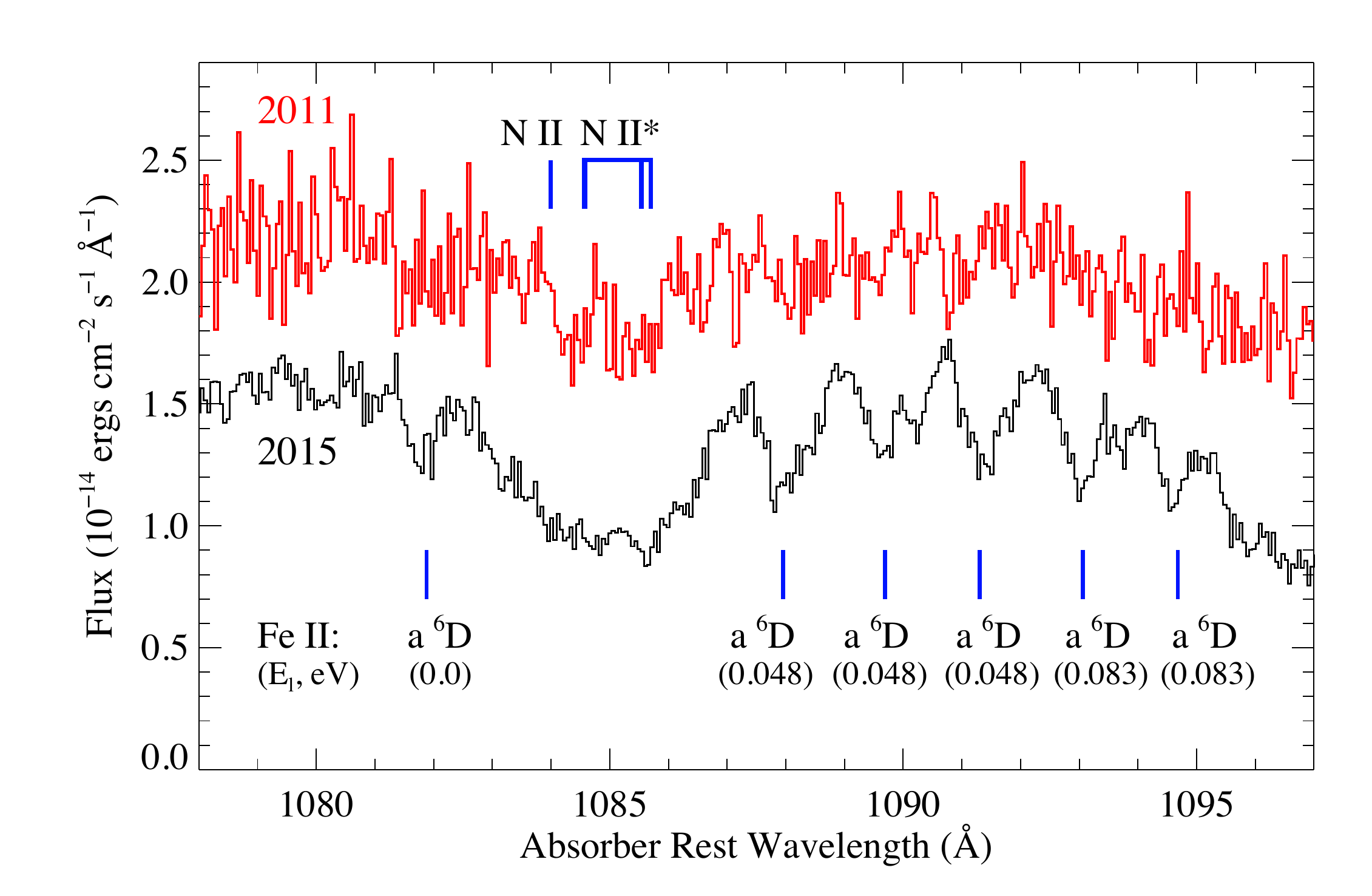}
\vspace{-13pt}
\caption{HST-COS spectra of PG~1411+442 showing the emergence of narrow \feii\ and \feii* absorption lines in 2015 (black curve) compared to 2011 (red curve). The positions and lower-state energies of the \feii\ and feii* lines (all arising from the ground $a\,^6D$ term) are marked across the bottom. A much broader mini-BAL feature due to \nii\ and \nii* is present in both spectra but broader and deeper in 2015.}
\end{figure}

Figure 1 plots the HST-COS spectra across a limited wavelength range that clearly shows the emergence of narrow transient absorption lines in 2015 compared to 2011. The particular narrow lines labeled in this figure are \feii\ and \feii* arising from the ground $a^6D$ multiplet of Fe$^+$ (see Section 5.1). These lines, as measured in 2015, were critical to our initial identification of the transient absorption-line system because they are relatively strong and free of blends. 

Figures 2 and 3 show larger portions of the HST-COS spectra densely labeled at the wavelengths of lines detected or predicted to be significantly present in the transient outflow system (Section 5.1). The spectra obtained in 2011 and 2015 both exhibit the mini-BALs that were the original focus of our study (Section 1). This mini-BAL system includes common outflow lines like \civ \lam 1548,1551 , \nv \lam 1239,1243, and \lya\ plus rare features like \pv\ \lam 1118,1128, low-ionisation lines of \cii\ \lam 1334, \siii\ \lam 1260, and \siii \lam 1527,  and excited-state lines \siv * \lam 1073, \ciii * \lam 1175, \siii * \lam 1265, and \siii * \lam 1533 that can provide density and location constraints. The mini-BALs measured in 2011 have velocity shifts of roughly v$\,\sim -1900$ \kms\ and a range of profile widths up to FWHM~$\sim 1300$ \kms\ in \civ\ and \nv . They become broader and slightly deeper in 2015. Two key results from our analysis of the mini-BALs in HTRV2 are that 1) the low-abundance \pv\ lines require total hydrogen column densities of $\log N_{\rm H} ({\rm cm}^{-2}) \gtrsim 22$ (for roughly solar P/H abundances), and 2) the excited-state lines of \siv * and \ciii * indicate that densities of $\log n_{\rm H} ({\rm cm}^{-3}) \gtrsim 5$ are present.  

\begin{figure*}
\includegraphics[scale=0.79]{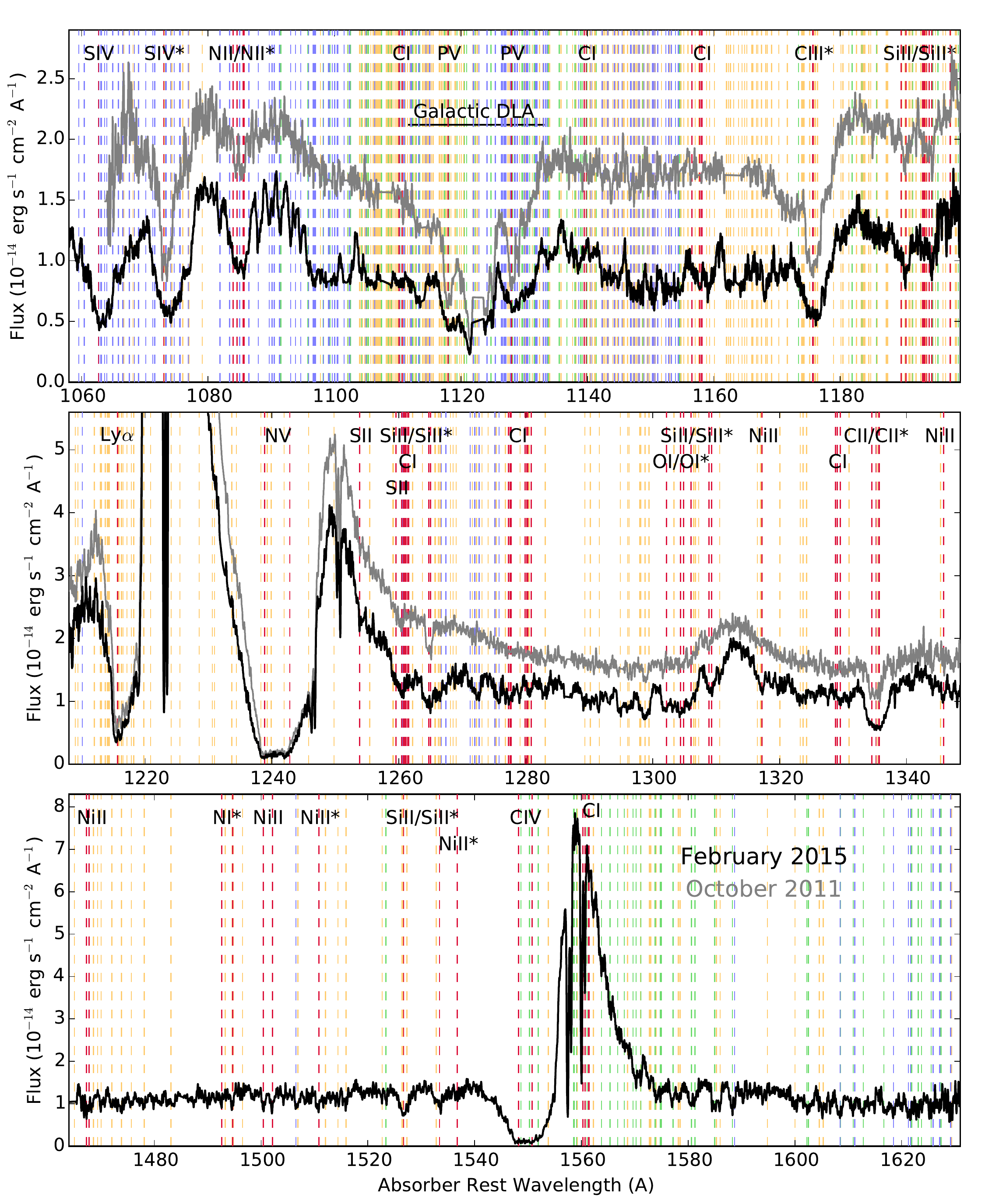}
\vspace{-15pt}
\caption{HST-COS spectra of PG1411+442 from 2015 (black) and 2011 (grey) in the absorber frame at $z_a = 0.08292$. Peculiar 15 to 45 percent ripples caused by the transient absorber became prominent in 2015. Dashed vertical lines mark absorption lines expected to be present. The red dashed lines indicate non-\feii\ features labeled across the top of each panel. All other dashed vertical lines are \feii\ transitions: blue = arising from the ground $a\,^6$D multiplet, green = from the next higher energy multiplet $a\,^4$F, orange = from other excited states up to 4.0 eV (see Section 5.1). Apparent narrow emission features in the 2015 spectrum, e.g., near 1103 \AA , 1146 \AA , 1161 \AA , 1301 \AA , and 1306 \AA , are small gaps in the rich absorption-line spectrum. Broad Galactic damped \lya\ (DLA) absorption is marked at $\sim$1112 to $\sim$1135 \AA . Other Galactic lines are removed and the spectra are smoothed for clarity.}
\end{figure*}

\begin{figure*}
\includegraphics[scale=0.77]{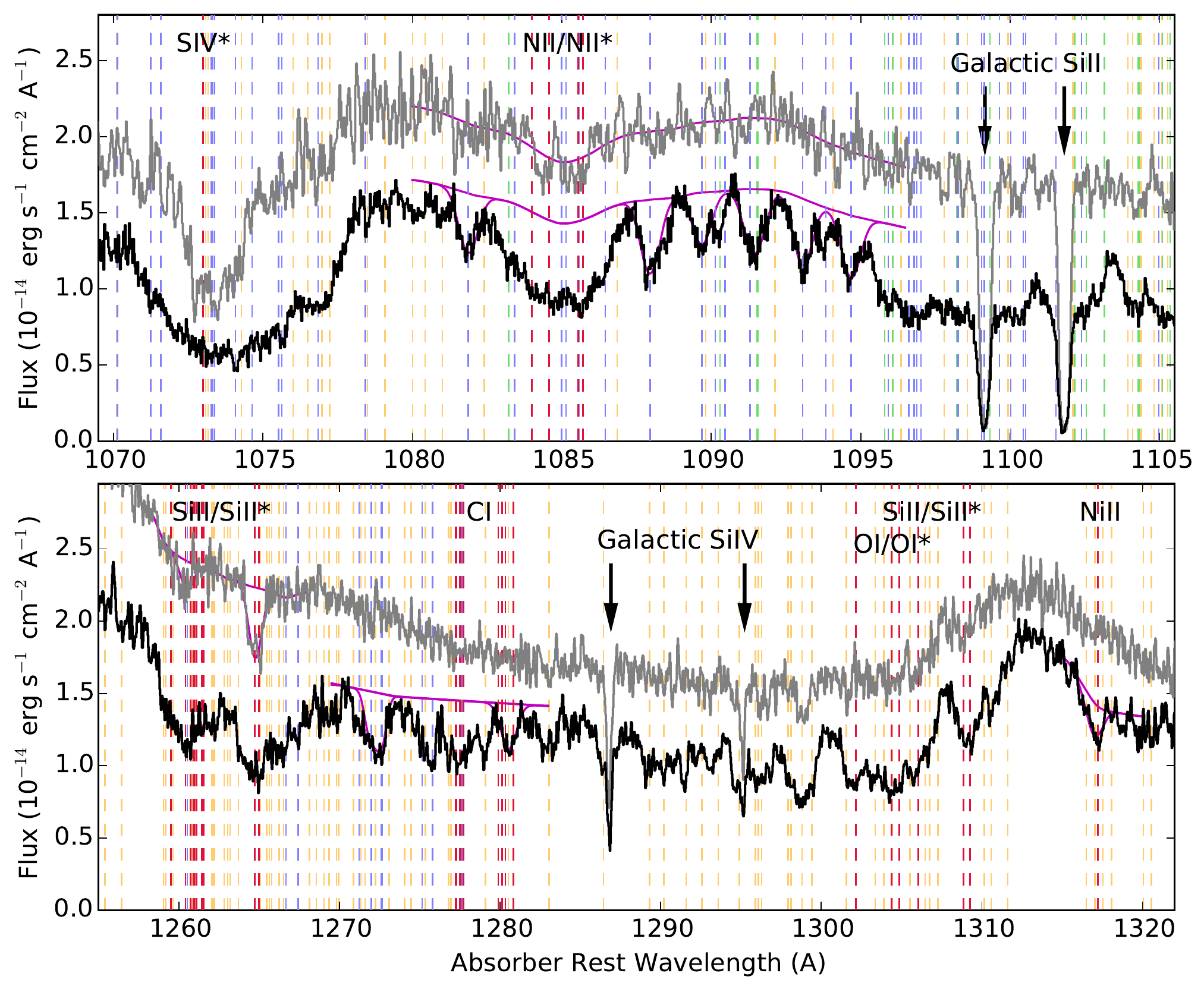}
\vspace{-8pt}
\caption{HST-COS spectra of PG 1411+442 showing two wavelength windows where lines from the transient absorber are clearly evident in 2015 (black spectrum) compared to the smoother 2011 spectrum (grey). The top panel includes the broad OI 1304 \AA\ broad emission line (shown here at $\sim$1313 \AA ) distorted by transient absorption lines in 2015. The bottom panel shows a cluster of distinct \feii\ and \feii* absorption lines in the transient system at wavelengths 1082--1095 \AA. Magenta curves show our fits to the continuum and selected lines in the transient system, as described in Section 5.2. Galactic absorption lines are marked by arrows above the spectra. All other notations are the same as Figure 2. }
\end{figure*}

Figures 2 and 3 also show the startling emergence in 2015 of \textit{many} weak narrow absorption lines we call the transient system. The large number of blended lines in the transient system gives the appearance of ripples in the spectrum with amplitudes of $\sim$15 to $\sim$45 percent. This peculiar ripple appearance led us initially to perform numerous checks to determine if the features are real in the PG~1411+442 spectrum or some type of artifacts or anomalies\footnote{These checks of the HST-COS data included 1) confirmation that the same ripple features appear in overlapping wavelength regions measured with two settings of the G130M grating (with the lines on different parts of the detector during the same visit), 2) the absence of similar features in COS spectra of other targets measured immediately before and after our 2015 PG1411+442 observations, and 3) reprocessing the spectra to rule out data reduction and detector anomalies.}. However, the reality of the ripples caused by absorption lines in PG~1411+442 is now confirmed directly subsequent HST-COS observations in 2016 and 2017 that also show the ripples (HTRV2), and by our visible Gemini-GMOS measurements showing similar peculiar absorption lines $\sim$4 months after the 2015 HST-COS observation (Table 1). 

Figure 4 plots the Gemini-GMOS spectrum on a vertical flux scale suitable for showing the emission lines. The weak transient absorption lines barely noticeable (with the strongest marked by blue arrows). Figure 5 zooms in on portions of this spectrum to show the transient lines more clearly. We derived a redshift for the transient system of $z_a \approx 0.08292\pm 0.00004$, corresponding to an outflow velocity of v~$\,= -1900\pm 11$ \kms , from the average line-center redshift in the strongest unblended transient lines in the Gemini-GMOS spectrum. The Doppler widths of these lines are roughly $b\sim 100$ \kms\ corresponding to FWHM $\sim$ 170 \kms\ (see Table 2 and Section 5.2 below). 
 
\begin{figure*}
\includegraphics[scale=0.7]{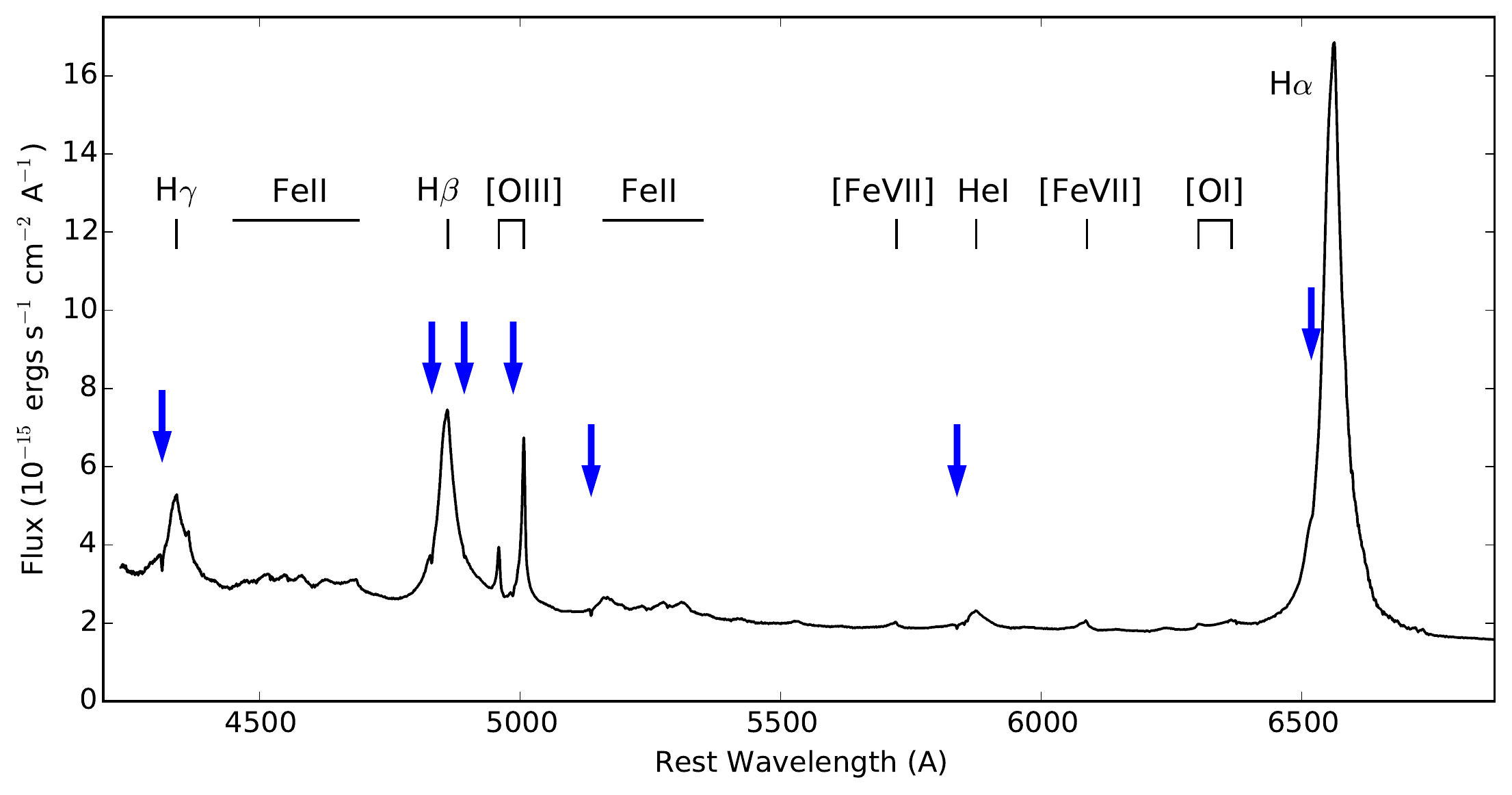}
\vspace{-8pt}
\caption{Gemini-GMOS spectrum of PG1411+442 from 2015. Prominent broad emission lines are labeled across the top. Some of the strongest absorption lines in the transient system are marked by blue arrows but are barely visible in this plot.}
\end{figure*}

\begin{figure*}
	\includegraphics[scale=0.53]{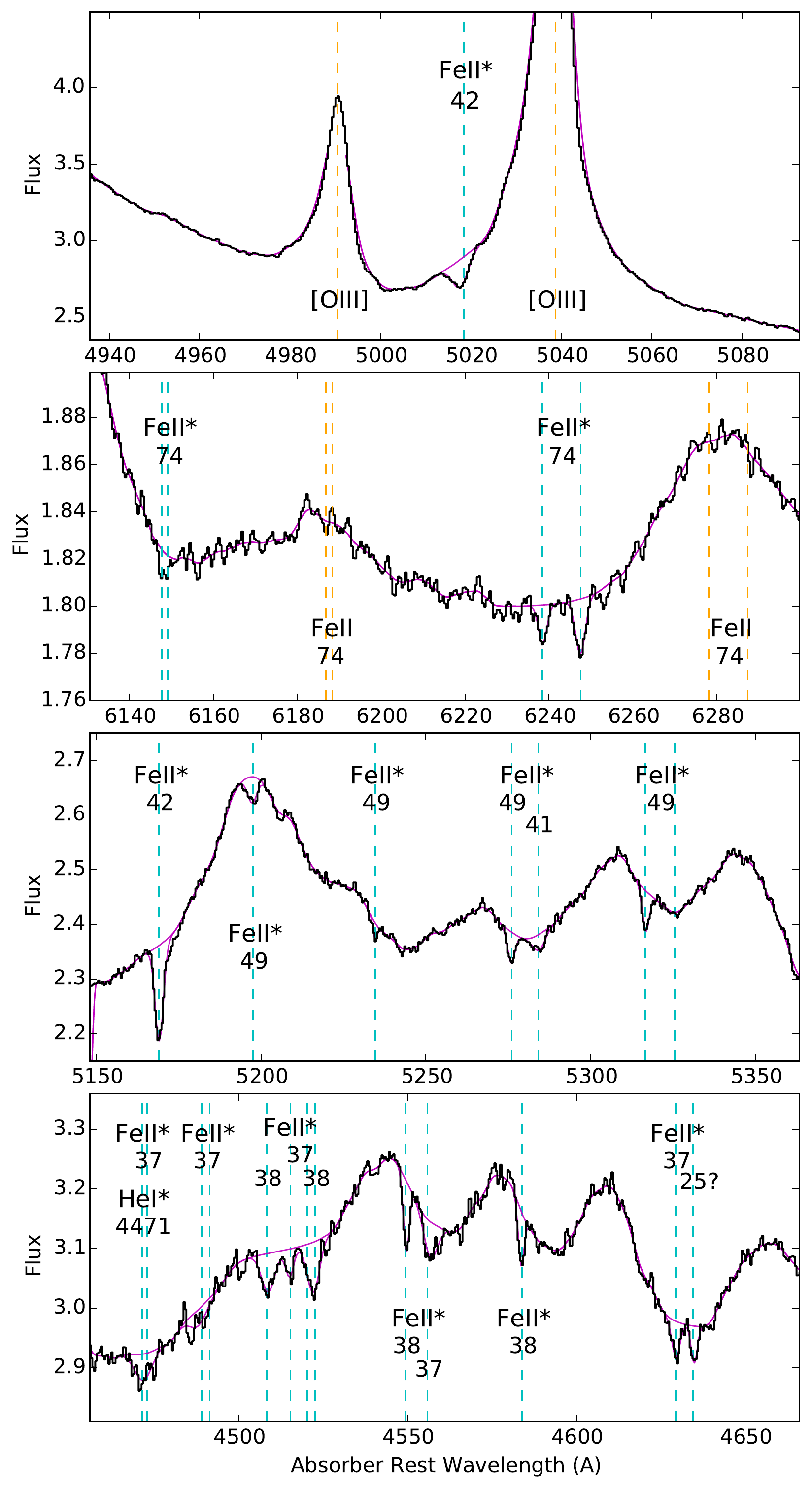}
	\includegraphics[scale=0.53]{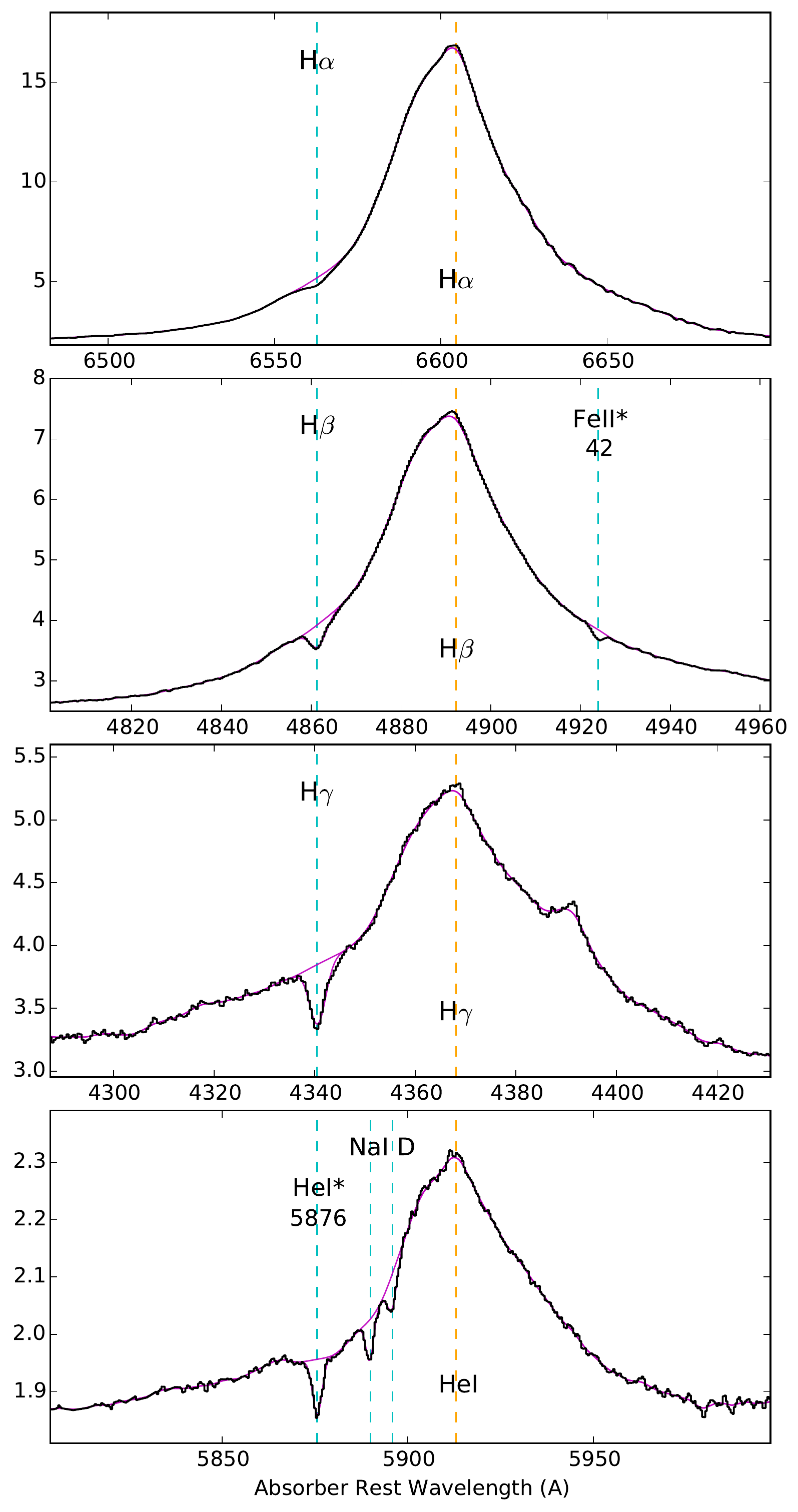}
\vspace{-5pt}
\caption{Segments of the 2015 Gemini/GMOS spectrum showing narrow absorption lines in the transient system. Wavelengths are plotted in the absorber frame. Narrow absorption in the transient system are marked by blue dashed vertical lines, while orange dashed vertical lines mark the wavelengths of some of the emission lines. Smooth magenta curves show the pseudo-continuum and fits to the absorption line profiles. The line fits are generally excellent and not clearly evident here under the observed spectra. The wavelength ranges in the right-hand panels span the same range in velocity to facilitate comparisons between the line profiles. The flux scale is the same as Figure 4. See Sections 5.1 and 5.2.}
\end{figure*}
 
The mini-BALs and the transient absorption lines appear to be physically related in the PG~1411+442 outflow based on similar centroid velocity shifts, narrow profiles in the low-ionization mini-BALs approaching the widths of the transient lines, and related line variabilities between 2011, 2015, and subsequent observations (HTRV2). Nonetheless, the transient lines are distinct enough in the spectrum and in their physical properties to warrant a separate analysis. They are exclusively narrower than the mini-BALs measured in the same epoch, and they emerged dramatically in 2015 (and faded in subsequent observations) while the mini-BALs remain present throughout all of our observations, and they require more extreme physical conditions (Section 5.4) that are not present in the mini-BAL gas. 

\section{Analysis of the Transient Absorber}

In this section, we discuss the transient line identifications and measurements, and we present an analysis for the absorber physical conditions as it appeared in our HST-COS and Gemini-GMOS spectra obtained in 2015. We do not expect line variability between these two 2015 observing epochs to affect our analysis because subsequent observations in 2016 (HTRV2) showed that the transient lines weakening only slightly from 2015, suggesting that the timescale for important variations was longer than 4 months. Also, our main conclusions about the outflow physical conditions rest primarily on well-measured lines in the Gemini-GMOS spectrum alone, which are corroborated by the extremely rich blended spectrum of lines measured with HST-COS in the UV. 

\subsection{Line Identifications}

We identify lines in the transient absorber using line wavelengths and atomic data from \cite{Verner96} and \cite{Wiese09} and the online databases managed by the National Institute of Standards and Technology\footnote{https://physics.nist.gov/PhysRefData/ASD/lines\_form.html} (NIST) and by Robert L. Kurucz\footnote{http://kurucz.harvard.edu/linelists.html} (version 08 October 2017). We also use laboratory measurements of UV \feii\ emission lines by \cite{Nave13}, and we refer to published line lists for the luminous blue variable star, $\eta$ Carinae \citep[][]{Nielsen05, Zethson12}, which has rich spectrum of low-ionisation emission/absorption lines that form in circumstellar ejecta \citep{Hamann12b}. 

The bottom portion of Table 2 lists every absorption line detected in the Gemini-GMOS spectrum. The table uses common names for the \nai~D and \hi\ Balmer lines, multiplet numbers from \cite{Moore46} and \cite{Moore62} when available, or lower-state term designations in other cases. Lines are listed with the `*' notation only if all lines in the group or mulitplet shown arise from excited states. The number of absorption lines detected in the visible is remarkable. Most of them are shown in Figure 5. To our knowledge, \hei* \lam 5876 and some of the excited-state \feii* lines have not been measured previously in any AGN outflow. The only previous report of visible \feii* lines in an AGN spectrum was by \cite{Shi16}. The \hi* Balmer and \nai~D absorption lines are rare, but they have been measured in several AGN outflows including the Seyfert 1 galaxy NGC~4151 \citep{Anderson71, Hutchings02} and some FeLoBAL quasars \citep{Boroson92, Rupke02, Rupke05, Hall02, Hall07, Ji15, Zhang15, Shi16, Sun17}. The Balmer lines were also measured in the transient absorber in PG~1411+442 by \cite{Shi17}

\begin{table*}
	\centering
	\caption{Absorption line data. Top portion: A small sample of lines in the 2015 HST-COS data plus \siii\ UV4 measured in the 2011. Bottom:  All lines detected in the 2015 Gemini/GMOS spectrum. The columns are: ID = line identification; $\lambda_r$ = rest wavelength (in vacuum for $\lambda_r < 2000$ \AA\ or air elsewhere); $E_l$ = lower state energy of the transition; $\lambda_{obs}$ = observed centroid wavelength; REW = rest equivalent width; v = velocity shift of the line centroid from $z_e = 0.08982$; $b$ = Doppler width; $\tau_o$ = apparent line-center optical depth. Notes in the last column: ID? = line ID is uncertain; bl = line appears blended (the blended line is indicated if it is also in this table); unc = fit data are unusually uncertain due to noise, blending, or continuum placement; tel = the line is contaminated by telluric absorption; $\tau_o^{c}$ = revised $\tau_o$ for Balmer lines to correct for broad emission lines filling in the troughs (see Section 5.4.2).}
	\begin{tabular}{lccccccccl} 
\hline
ID & $\lambda_r$ & $E_l$ & $\log(gf)$ & $\lambda_{obs}$ & REW & v & $b$& $\tau_o$ & Notes \\
&  (\AA ) & (eV) & & (\AA ) & (\AA ) & (\kms ) & (\kms ) \\
\hline

\ci\ ~~UV5& 1280.36 & 0.00 & --0.72 & 1386.50 &  0.288$\pm$0.041 &  $-$1863$\pm$11 &  154$\pm$16 &  0.247& unc, bl--\ci *\cr
\nti *~UV4 & 1494.68 & 2.36 & --0.63 & 1618.59 &  0.262$\pm$0.066 &  $-$1931$\pm$12 &  89$\pm$16 &  0.333\vspace{3pt} \cr

\siii\ ~UV2& 1526.71 & 0.00& --0.57& 1653.27 &  1.332$\pm$0.113 &  $-$1881$\pm$11 &  287$\pm$15 &  0.515 & bl\cr
& 1533.43 & 0.036 & --0.27 & 1660.55 &  0.898$\pm$0.086 &  $-$1881$\pm$11 &  287$\pm$15 &  0.347\vspace{3pt}  \cr

\siii\ ~UV4& 1260.42 & 0.00& 0.39& 1364.91 &  0.084$\pm$0.014 &  --1889$\pm$7 &  130$\pm$9 &  0.087& 2011, unc \cr
& 1264.80 & 0.036& 0.68& 1369.59 &  0.239$\pm$0.025 &  --1889 $\pm$7 &  130$\pm$9 &  0.247& 2011\vspace{3pt} \cr

\feii\ ~$a\,^6{\rm D}$& 1081.88 & 0.00 & --0.71 & 1171.56 &  0.207$\pm$0.015 &  --1899$\pm$6 &  117$\pm$4 &  0.274 \cr
& 1087.95 & 0.048 & --0.95 & 1178.17 &  0.276$\pm$0.016 &  --1899$\pm$6 &  117$\pm$4 &  0.365 \cr
& 1089.69 & 0.048 & --1.08 & 1180.04 &  0.178$\pm$0.013 &  --1899$\pm$6 &  117$\pm$4 &  0.235 \cr
& 1091.30 & 0.048& --2.60& 1181.79 &  0.109$\pm$0.020 &  --1899$\pm$6 &  117$\pm$4 &  0.144 & bl -1091.55\cr
& 1093.06 & 0.083 & --1.18 & 1183.69 &  0.243$\pm$0.015 &  --1899$\pm$6 &  117$\pm$4 &  0.321 \cr
& 1094.68 & 0.083 & --1.04 & 1185.45 &  0.243$\pm$0.015 &  --1899$\pm$6 &  117$\pm$4 &  0.322\cr
& 1271.99 & 0.083 & $-$1.06 & 1377.43 &  0.182$\pm$0.039 &  --1916$\pm$18 &  105 &  0.230 & bl--1272.64\cr
& 1272.63 & 0.083& $-$1.25 & 1378.13 &  0.225$\pm$0.038 &  --1916$\pm$18 &  105 &  0.285 & bl--1271.99 \vspace{3pt} \cr

\feii* ~$a\,^4{\rm F}$& 1091.55 & 0.23 & --1.27 & 1182.07 &  0.136$\pm$0.020 &  --1899$\pm$6 &  117$\pm$4 &  0.180 & bl--1091.30\vspace{3pt} \cr

\feii* ~UV44& 1580.63 & 0.30 & --0.66 & 1711.66 &  0.511$\pm$0.090 &  --1898$\pm$12 &  110$\pm$16 &  0.497 \cr
& 1584.95 & 0.35 & --0.75 & 1716.34 &  0.532$\pm$0.095 &  --1898$\pm$12 &  110$\pm$16 &  0.518 & bl \cr
& 1588.29 & 0.39 & --0.96 & 1719.96 &  0.459$\pm$0.097 &  --1898$\pm$12 &  110$\pm$16 &  0.447\vspace{3pt}  \cr

\feii\ ~UV8& 1608.45 & 0.00 & --0.23 & 1741.79 &  0.452$\pm$0.102 &  --1918$\pm$14 &  101$\pm$18 &  0.470\vspace{3pt}  \cr

\nkii\ ~$a^2{\rm D}$& 1317.22 & 0.00 & --0.05 & 1426.42 &  0.206$\pm$0.041 &  --1956$\pm$16 &  152$\pm$22 &  0.174& unc, bl?\cr
& 1345.88 & 0.00 & --1.41 & 1457.45 &  0.159$\pm$0.044 &  --1938$\pm$13 &  88$\pm$18 &  0.228 \cr
& 1500.43 & 0.19 & --0.71 & 1624.82 &  0.311$\pm$0.060 &  $-$1900$\pm$8 &  89$\pm$10 &  0.395 \cr
& 1502.05 & 0.00 & --1.10 & 1626.57 &  0.301$\pm$0.059 &  $-$1900$\pm$8 &  89$\pm$10 &  0.382\vspace{-2pt} \cr

\multicolumn{10}{c}{\dotfill}\vspace{2pt} \cr

H$\gamma$&  4340.47 & 10.2& --0.45 & 4700.50 &  0.521$\pm$0.024 &  --1889$\pm$4 &  154$\pm$5 &  0.132 & $\tau_o^{c} \approx 0.16$\cr
H$\beta$& 4861.34 & 10.2& --0.02 & 5264.64 &  0.431$\pm$0.017 &  --1884$\pm$3 &  150$\pm$4 &  0.100 & $\tau_o^{c} \approx 0.16$\cr
H$\alpha$& 6562.83 & 10.2& +0.71 & 7107.15 &  0.602$\pm$0.011 &  --1893$\pm$2 &  213$\pm$3 &  0.073 & $\tau_o^{c} \approx 0.22$ \vspace{3pt}\cr

\hei* $\,2p\,^3{\rm P}^o$ & 4471.48 & 21.0& +0.04 & 4842.54 &  0.094$\pm$0.020 &  --1880$\pm$26 &  235$\pm$37 &  0.015& bl? \cr
& 5876.62 & 21.0& +0.74 & 6362.87 &  0.159$\pm$0.006 &  --1896$\pm$2 &  92$\pm$3 &  0.050\vspace{3pt} \cr

\nai~D& 5889.95 & 0.00& +0.11 & 6378.16 &  0.094$\pm$0.005 &  --1904$\pm$2 &  78$\pm$3 &  0.035 \cr
& 5895.92 & 0.00& --0.19 & 6384.74 &  0.085$\pm$0.004 &  --1904$\pm$2 &  78$\pm$3 &  0.031\vspace{3pt} \cr

\feii* ~37,38 & 4491.40 & 2.84& --2.71 & 4860.26 &  0.057$\pm$0.020 &  --2118$\pm$43 &  227$\pm$61 &  0.010 & bl\cr
& 4508.28 & 2.86 & --2.33 & 4882.65 &  0.104$\pm$0.012 &  --1865$\pm$11 &  184$\pm$16 &  0.021 \cr
& 4515.33 & 2.85 & --2.50 & 4889.49 &  0.046$\pm$0.011 &  --1914$\pm$15 &  114$\pm$21 &  0.015 \cr
& 4522.63 & 2.83 & --2.14 & 4897.09 &  0.125$\pm$0.020 &  --1773$\pm$15 &  171$\pm$21 &  0.027& bl \cr
& 4549.47 & 2.83 & --1.97 & 4926.90 &  0.080$\pm$0.009 &  --1887$\pm$4 &  78$\pm$6 &  0.038 \cr
& 4555.89 & 2.83 & --2.35 & 4934.78 &  0.093$\pm$0.035 &  --1830$\pm$35 &  176$\pm$50 &  0.020& bl \cr
& 4583.83 & 2.81 & --1.81 & 4963.71 & 0.061$\pm$0.008 &  --1911$\pm$5 &  76$\pm$7 &  0.030  \cr
& 4629.33 & 2.81 & --2.31 & 5013.34 &  0.060$\pm$0.020 &  --1890$\pm$19 &  109$\pm$27 &  0.020\vspace{3pt} \cr

\feii* ~25& 4634.61 & 2.58 & --5.58 & 5019.20 &  0.049$\pm$0.019 &  --1881$\pm$17 &  83$\pm$24 &  0.022& ID?\vspace{3pt} \cr

\feii* ~42& 4923.92 & 2.89& --1.56 & 5332.15 &  0.131$\pm$0.006 &  --1901$\pm$2 &  103$\pm$3 &  0.044 \cr
& 5018.44 & 2.89 & --1.40 & 5433.75 &  0.276$\pm$0.015 &  --1943$\pm$4 &  142$\pm$6 &  0.065 \cr
& 5169.03 & 2.89 & --1.30 & 5597.80 &  0.242$\pm$0.013 &  --1890$\pm$3 &  103$\pm$4 &  0.075\vspace{3pt} \cr

\feii* ~49& 5197.57 & 3.23 & --2.23 & 5628.29 &  0.065$\pm$0.008 &  --1912$\pm$8 &  118$\pm$11 &  0.018 & unc \cr
& 5234.62 & 3.22 & --2.14 & 5668.69 &  0.019$\pm$0.005 &  --1897$\pm$7 &  47$\pm$9 &  0.013 \cr
& 5276.00 & 3.20 & --2.06 & 5713.36 &  0.065$\pm$0.008 &  --1905$\pm$6 &  87$\pm$8 &  0.024\cr
& 5316.61 & 3.15 & --1.94 & 5757.57 &  0.079$\pm$0.006 &  --1893$\pm$3 &  82$\pm$5 &  0.030\vspace{3pt} \cr

\feii* ~41& 5284.09 & 2.89 & --3.41 & 5722.52 &  0.054$\pm$0.023 &  --1884$\pm$33 &  143$\pm$47 &  0.012& unc, ID?\vspace{3pt} \cr

\feii* ~74& 6148.5~~ & 3.89 & --2.73 & 6658.3~~ & 0.018$\pm$0.004 & --- & --- & ---& bl, unc\cr
& 6238.39 & 3.89 & --2.64 & 6755.79 &  0.023$\pm$0.004 &  --1892$\pm$6 &  69$\pm$8 &  0.009 \cr
& 6247.56 & 3.89 & --2.34 & 6765.52 &  0.039$\pm$0.005 &  --1902$\pm$5 &  77$\pm$7 &  0.014 \cr
& 6416.92 & 3.89 & --2.75 & 6948.12 &  0.070$\pm$0.010 &  --1937$\pm$5 &  60$\pm$6 &  0.031& unc, tel \cr
& 6456.38 & 3.90 & --2.09 & 6991.28 &  0.056$\pm$0.008 &  --1918$\pm$6 &  87$\pm$9 &  0.017 & unc, tel \cr
\hline
\end{tabular}
\end{table*}

Line identifications in the UV are much more difficult due to line blending. We will show below that there are more than a thousand absorption lines present across the wavelengths plotted in Figures 2 and 3. The top portion of Table 2 lists a small subset of these features that appear relatively free of blends. This includes two pairs of \siii\ and \siii * lines that have narrow profiles near the ambiguous boundary between the mini-BALs and the transient system. We will argue in Section 5.4 (also HTRV2) that these lines belong mainly to the mini-BAL outflow. 

We identify transient lines in the 2015 UV spectrum involves marking the positions of plausible absorption lines in plots like Figures 2 and 3 and then checking for coincidences with actual features in the observed spectrum. This procedure is iterative and inexact, but we are guided by the lines securely identified in Table 2 and Figure 5, and by our theoretical analysis in Sections 5.3 and 5.4 that makes specific line strength predictions. Our goal is to create a \textit{conservative} line list and labelling scheme for Figures 2 and 3 that includes all of the strongest lines that are present or likely to be present in the transient system, while excluding large numbers of weaker lines (that might also contribute to the spectrum). The dashed vertical lines in Figures 2 and 3 show the results. Careful inspection of these figures shows that the labels do account for most of the observed lines/ripples in the transient system. The line labeling scheme is as follows: 

The red dashed vertical lines in Figures 2 and 3 mark all non-\feii\ transitions from low energy states in abundant ions that are likely to be present in the transient system. The lines thus labeled arise from ground multiplets at energies $<$0.25 eV with transition strengths $\log (gf\lambda ) > 1.45$ (where $g$ is the statistical weight, $f$ is the oscillator strength, and the wavelength $\lambda$ is in \AA , with data from the Kurucz database). Exceptions to this also labeled are \ciii * \lam 1175 and \nti * 1492.63 \AA\ and 1494.68 \AA, which arise from metastable states at much higher energies of 6.4 eV and 2.4 eV, respectively. \ciii * \lam 1175 is a blend of closely-spaced lines that we label for convenience as a single transition. The \nti * identifications are somewhat tentative because only the 1494.68 \AA\ line is  securely measured (Figure 2, Table 2). However, their identifications are supported by detections of these lines in absorption in $\eta$ Carinae \citep{Nielsen05}, which has an absorption-line environment similar to the transient absorber in PG~1411+442 \citep[see also][]{Hamann12b}. 

The \ci\ lines marked in red in Figures 2 and 3 are also somewhat tentative because they are weak and blended. There are actually many more \ci\ and \ci * lines within our wavelengths coverage, but none are clearly present and they are excluded from our labeling scheme by the $gf\lambda$ lower limit. The combination of weak absorption (caused in part by partial covering) and extensive line blending makes features like this very difficult to assess in the spectrum (see further discussion below). The most secure \ci /\ci * identifications are the blended multiplets at $\sim$1277.5 \AA\ and $\sim$1280.4 \AA . Not marked in Figure 2 but included in our line list is a set of \feiii\ UV1 lines between $\sim$1122 \AA\ and $\sim$1132 \AA . These lines are likely to be present in transient system but they are not apparent in the observed spectrum due to blending with numerous \feii\ and \feii* lines and the strong \pv\ \lam 1118,1128 mini-BALs. 

All of the other dashed vertical lines in blue, green, and orange in Figures 2 and 3 mark \feii\ and \feii* transitions, which are by far the most numerous to the 2015 spectrum. The blue and green dashed lines mark lines arising from the lowest two energy terms, $a\,^6$D and $a\,^4$F, respectively, while the orange dashed lines mark transitions from higher energy states up to 4.0 eV. All of the \feii\ and \feii* transitions marked in Figures 2 and 3 satisfy a minimum strength index that we define as $P \equiv \log[(gf\lambda )e^{-E_l/kT}]$, where $\lambda$ is the rest wavelength in \AA , $E_l \leq 4.0$ eV is the lower-state energy, and $T=10,000$ K is the excitation temperature (using atomic data from the Kurucz database). This prescription is guided by our analysis in Sections 5.3 and 5.4 indicating that the Fe$^+$ level populations are roughly in thermal equilibrium at high densities {\citep[also][]{Wampler95}. We also add to our line list a small number of strong laboratory \feii\ and \feii* lines \citep[from][]{Nave13} that appear to be present in the PG~1411+442 spectrum but they are not in our automated search (perhaps due to errors in the theoretical oscillator strengths). 

We experimented with many different values of $T$ and $P$ to create an \feii\ and \feii* line list that marks most of the observed features without listing every weak line that might contribute. The well-measured lines in the visible spectrum provide a good test. The combination $T=10,000$ K and $P\geq -0.3$ successfully predicts all of the \feii* lines in multiplets 37, 38, 42, and 49 but not the four weakest lines in the highest-energy multiplet 74 (at 3.89 eV, see Table 2). We manually add these lines to our list, but we do not raise $T$ or lower $P$ to capture them in our automated search because $T=10,000$ K is already hotter than the Fe$^+$ zone in our \cloudy\ models (below) and lowering $P$ by the required amount would add more than a thousand other weak \feii\ and \feii* lines to our list, mostly in the UV. Figure 2 uses a higher threshold of $P\geq 0.2$ to limit the density of line labels in that plot for convenience, but zooming in on selected regions shows that this $P$ threshold is too high for the UV (and much too high for the visible). Figure 3 and our final line list use $T=10,000$ K and $P\geq -0.3$. 

Inspection of Figures 2 and 3 shows that our adopted labeling scheme marks all of the distinct lines listed in Table 2 plus many others that are weak or more blended. It also indicates that many UV \feii\ and \feii* lines blend together to suppress the measured flux across wide wavelength bands in the observed 2015 spectrum. Small gaps between these absorption bands give the (false) appearance of narrow emission lines, e.g., near 1103 \AA , 1137 \AA , 1146 \AA , 1161 \AA , 1301 \AA , and 1306 \AA . 

Altogether, our line identification scheme indicates that $\geq$1632 metal lines (not counting H and He) contribute significantly to the transient spectrum at UV wavelengths from 940 \AA\ to 1630 \AA\ (quasar frame), and that $\geq$2990 metal lines contribute from 940 \AA\ to $\sim$6500 \AA . These numbers are lower limits because our list is conservative and many weak lines that probably also contribute are not included. The final line list is available upon request from the authors.
%Finally, we searched for but do not find molecular lines of H$_2$ or CO in the transient absorber. In particular, significant amounts of H$_2$ should produce distinctive broad absorption bands at XXX, XXX, and XXX \AA\ \citep{XXX}, which are not apparent in the observed spectrum. 

\subsection{Line Measurements}

We fit only the select group of absorption lines listed in Table 2 because they appear relatively free of blends and/or they are important for our analysis below. The fits use Gaussian optical depth profiles of the form
\begin{equation}
\tau_{\textrm{v}} \ = \ \tau_{o}\, e^{-{\textrm{v}}^2/b^2}
\end{equation}
where $\tau_o$ is the line-center optical depth given by  
\begin{equation}
\tau_o\ =\ {{\sqrt{\pi}\, e^2}\over{m_e\, c}}\,{{N_l\,f\lambda_r}\over{b}}
\end{equation}
and  v is the velocity in the quasar frame, $b$ is the width Doppler parameter, $\lambda_r$ is the laboratory rest wavelength, and $N_l$ is the ion column density in the lower energy state, $l$, of the transition. We will argue in Section 5.4 that the transient absorber only partially covers the quasar emission source. A simple way to characterize the observed line intensities with partial covering is
\begin{equation}
{{I_{\textrm{v}}}\over{I_{\textrm{v}}^c}} \ = \ (1-C_{\textrm{v}}) + C_{\textrm{v}}\,e^{-\tau_{\textrm{v}}}
\end{equation}
where $I_{\textrm{v}}^c$ is the emitted (unabsorbed) intensity, $C_{\textrm{v}}$ is the covering fraction with values $0<C_{\textrm{v}}\leq 1$, and $\tau_{\textrm{v}}$ is the line optical depth at velocity shift v \citep{Hamann97, Barlow97}. This expression assumes implicitly that the absorber is homogeneous in front of a uniform emission source. However, an important feature of the transient absorber is that it is spatially inhomogeneous, presenting a range of column densities and optical depths across the emission source (Sections 5.4 and 5.6). Part of the emission source can be covered by gas that is optically thick in a given line while other parts are covered by optically thin material. In this situation, the covering fractions $C_{\textrm{v}}$ in Equation 3 correspond roughly to the fraction of the emission source covered by material with $\tau_{\textrm{v}} \gtrsim 1$ \citep{deKool02, Hamann04}. 

In principle, we could determine both $\tau_{\textrm{v}}$ and $C_{\textrm{v}}$ by solving Equation 3 simultaneously for two or more lines in multiplets with known optical depth ratios \citep[][]{Hamann97, Hamann11, Ganguly99, Gabel05, Arav05, Arav08}. However, this can lead to poor results for cases like PG~1411+442 with spatially inhomogeneous absorption because the measured lines can have optical depth-dependent strengths (even if they are very saturated) owing to optical depth-dependent covering fractions \citep{Hamann01, Hamann04, Arav08, Borguet12, Moravec17}. This is the main feature of inhomogeneous partial covering and it can lead to severe underestimates of the column densities. A related problem is that single $\tau_{\textrm{v}}$ values derived for inhomogeneous absorbers have ambiguous meanings. If the absorber spans a range of optical depths from thick to thin across the emission source, then the observed line ratios will tend to indicate intermediate $\tau_{\textrm{v}}$ regardless of the true range, with the derived values depending on the spatial \textit{gradients} in $\tau_{\textrm{v}}$ across the absorber. Outflow structures with very large column densities can go completely undetected in observed line troughs if they are more compact (with smaller covering fractions) than other outflow material. Another serious problem for measurements of the transient lines in PG~1411+442 is blending in the UV. 

We adopt a simple strategy to fit only select well-measured lines (Table 2) with the assumption of complete covering ($C_{\textrm{v}} = 1$). For simplicity, we also ignore any velocity dependence in the covering fractions. This strategy yields good measurements of the basic line properties and firm \textit{lower limits} on the line optical depths. We then examine the fit results for various lines, aided by photoionisation models, to derive constraints on the true covering fractions, optical depths, and column densities. 

For the Gemini-GMOS spectrum, we first construct a pseudo-continuum that includes the quasar continuum plus the broad emission lines. This is straightforward because the lines are sparse and narrow; we manually interpolate across the tops of the absorption lines and then smooth the results (to remove noise) using a simple boxcar function. The magenta curves in Figure 5 show this continuum and our fits to the absorption lines. 

The HST-COS spectrum from 2015 is more challenging because there are many blended absorption lines. For the cluster of \feii\ and \feii* lines at wavelengths $\sim$1082 to $\sim$1095 \AA , we adopt a continuum that is a smoothed version of the 2011 HST-COS spectrum scaled by a factor of 0.78 to touch the highest flux points between obvious absorption lines in the 2015 spectrum. This continuum and the resulting line fits are shown in the bottom panel of Figure 3. For other lines, we manually draw continua across small wavelength segments that encompass the absorption lines of interest. This is a subjective procedure with considerable uncertainties. The upper panel in Figure 3 shows several examples of these manually-drawn continua along with the line fitting results.

Figure 6 shows more examples of the line fits in normalized spectra on a velocity scale relative to the emission-line redshift, $z_e = 0.08983$. Table 2 lists the line parameters for all of the fitted lines, including every line detected in the visible Gemini-GMOS spectrum. It is important to emphasize that the {\it apparent} line-center optical depths in Table 2 are lower limits because they are derived with the assumption of unity covering fractions. 

\begin{figure*}
\begin{center}
\includegraphics[scale=0.55]{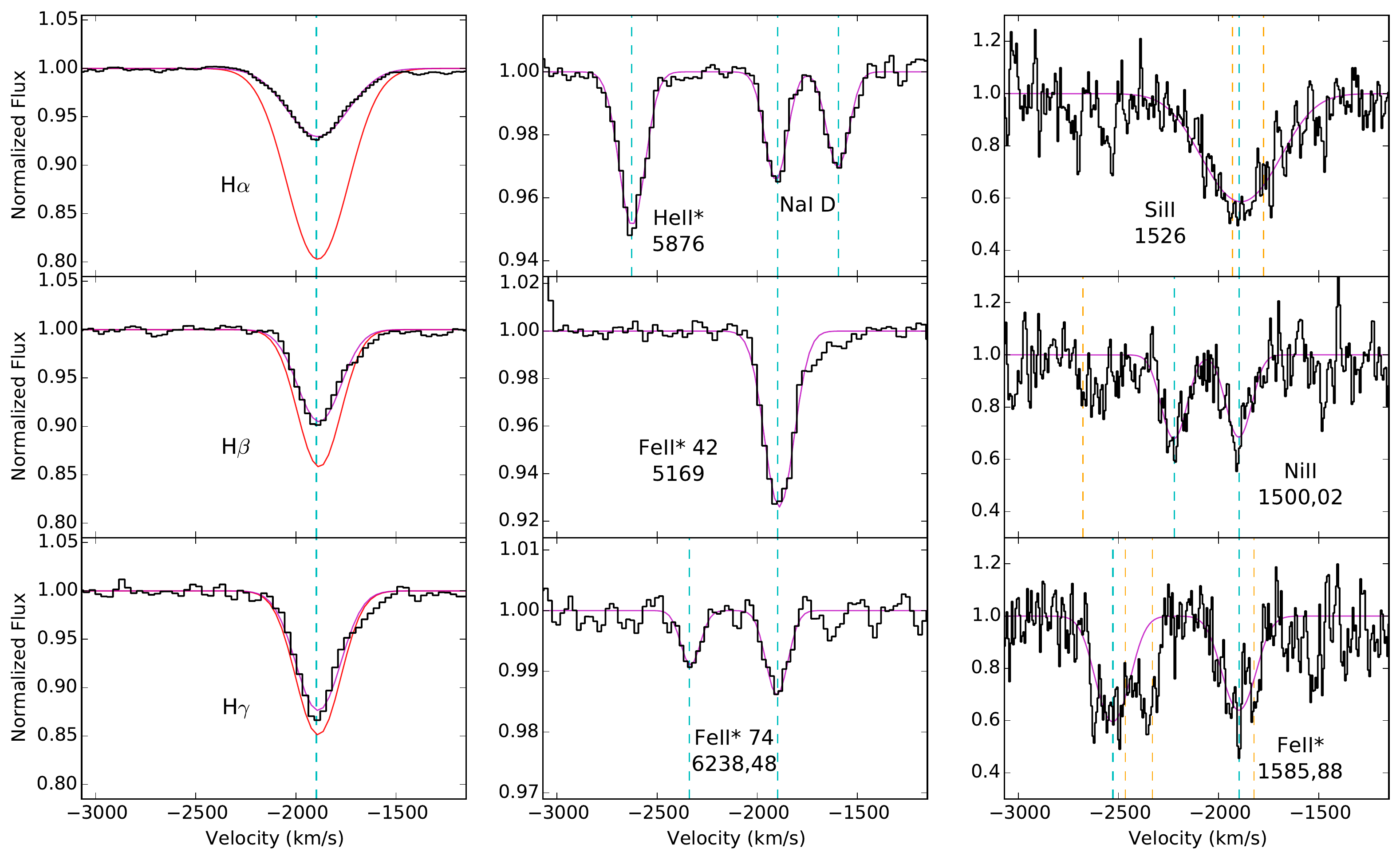}
\end{center}
\vspace{-12pt}
\caption{Normalized spectra from 2015 Gemini/GMOS and 2015 HST-COS showing representative line profiles in the transient absorber, as indicated in each panel. Smooth magenta curves show our fits to the continua and line profiles. Blue dashed vertical lines mark the nominal velocity shift of the absorber at v~$\approx -1900$ \kms . In panels with multiple fitted lines, the velocity scale applies only to the feature near v~$\approx -1900$ \kms . For the UV lines in the right-hand panels, the orange dashed lines marked \feii\ and \feii* lines should be present based on their strength index, $P$, in Section 5.1. Smooth red curves drawn for the Balmer lines show the fits corrected for broad emission-line flux filling in the absorption troughs (Section 5.4.2).}
\end{figure*}

\subsection{\cloudy\ Simulations}

We use the photoionisation and spectral synthesis code \cloudy\ \citep[version 17.00,][]{Ferland17} to predict absorption line strengths in different physical conditions. High densities leading to large excited-state populations are essential to produce the rich variety of lines seen in the transient system. For \hi , \hei , and \hi -like and \hei -like ions, we use default settings in \cloudy\ that solve the statistical equilibrium equations for states with principle quantum numbers up to $n\approx 25$ (with the lowest 6 $n$ states resolved into separate $l$ states). We use the full Fe$^+$ atom available in \cloudy , which includes the lowest 371 energy levels up to 11.6 eV. For all other ions, we again use default settings to calculate populations in the lowest 15 energy states in every ion. See the \cloudy\ manual for more details including the atomic datasets used. 

We adopt a standard quasar continuum shape that has power-law slopes across optical-to-UV and X-ray wavelengths of $\alpha_{uv}=-0.5$ and $\alpha_x=-0.9$, respectively, for $f_{\nu}\propto\nu^{\alpha}$. These power-law segments are joined smoothly in the far-UV by an exponential Wien function with temperature $T$ = 250,000 K. The relative strengths of the UV and X-ray spectral segments are scaled to yield a two-point power-law index from 2500 \AA\ to 2 keV of $\alpha_{ox} = -1.5$. This value of $\alpha_{ox}$ is based on observations of other quasars with luminosities similar to PG~1411+442 \citep{Strateva05,Steffen06}. We do not use the observed UV to X-ray flux ratio in PG~1411+442 directly because its X-ray spectrum appears to be affected by absorption and non-quasar sources of emission \citep{Teng10,Steenbrugge11}. The spectral shape overall is defined in \cloudy\ by the command: {\tt AGN T=250000K, a(ox)=-1.5, a(uv)=-0.5, a(x)=-0.9}. It yields bolometric corrections of $L \sim 4.0\, \lambda L_{\lambda}$(1500\AA ) or $L \sim 6.0\, \lambda L_{\lambda}$(5100\AA ).

The intensity of ionising radiation at the illuminated face of the clouds is set by the dimensionless ionisation parameter, 
\begin{equation}
U \ \equiv \ {{Q_H}\over{4\pi c\, R^2\, n_{\rm H}}}
\end{equation}
where $Q_H$ is the emitted luminosity of hydrogen-ionising photons (\#/s), $n_{\rm H}$ is the total hydrogen density, and $R$ is the radial distance from the quasar. For the spectral shape described above, the observed UV quasar flux (Section 3) yields this expression for the radial distance
\begin{equation}
R \ = \ 0.48\, \left({{\lambda L_\lambda (1500{\rm \AA })} \over{5\times 10^{44}\,{\rm ergs~s}^{-1}}}\right)^{1/2} \left({{10^8\,{\rm cm}^{-3}}\over{n_{\rm H}}}\right)^{1/2}
\left({{0.18}\over{U}}\right)^{1/2}~{\rm pc}~
\end{equation}

Large column densities and radiative shielding are important to produce the full range of low-ionisation absorption lines in the transient absorber. We do not consider scenarios where the shielding occurs in an external medium between the quasar and the outflow absorber because that simple picture is problematic \citep{Hamann13}. It has been proposed that a stationary or low-speed shielding medium at the base of the outflow is needed to moderate the outflow ionisations and increase the opacities for radiative driving in the outflow farther from the quasar \citep{Murray95, Murray97}. However, the shielding gas needs substantial bound-free opacities in the far-UV to shelter ions like \ovi , \nv , and \civ\ in the outflow. This would produce strong absorption lines of \ovi, \nv, and \civ\ in the shield (presumably at roughly zero velocity) that are not generally seen in real quasar outflows. 

For low-ionisation outflows like the transient absorber, the shield would need column densities large enough to support a \hii --\hi\  recombination front inside the shield. This would lead to many absorption lines of both high and low-ionisation species. Thus the shield as a separate entity is paradoxical because it needs to be self-shielding with physical conditions like the outflow gas it is intended to shield so it can have those properties. A related problem is that an effective shield with large bound-free opacities and strong UV absorption lines will also have the conditions needed for radiative driving. Therefore, the shielding occurs not in a separate medium at the base of the outflow, but in the outflow itself. 

The model clouds in our simulations are plane-parallel with solar abundances \citep{Asplund09}, constant density, and turbulent velocities consistent with the measured line widths, $b=100$ \kms . We assume the densities are constant everywhere inside the outflow clouds as a convenience for our discussions of density-dependent properties below. Constant pressure might be more realistic for the clumpy inhomogeneous absorber in PG~1411+442, but the density gradients needed to maintain constant pressure in the line-forming regions are too small to affect our main results. The model clouds are truncated at a maximum total column density of $\log N_{\rm H} ({\rm cm}^{-2}) = 23.6$ or at a depth where the electron temperature reaches $T\leq 4000$ K, whichever is reached first in the calculations. The upper limit on $N_{\rm H}$ ensures that the Thompson scattering optical depth is in all cases $\leq$26 percent (the absorber must be optically thin to Thompson scattering to produce a measurable absorption-line spectrum). The minimum temperature avoids extended cold/neutral regions that would produce many more absorption lines (including molecules) not detected in the transient absorber. The specific values of these limits are arbitrary, but the model clouds must be truncated somewhere and the limits we use ensure that all of the important lines in the transient spectrum are included in the calculations. 

Figures 7 and 8 and Table 3 summarize the main results from our \cloudy\ simulations. Figure 7 plots predicted line-center optical depths for selected lines as a function of $n_{\rm H}$ and $U$ (based on 322 separate calculations). Figure 8 shows the ionisation structure and some line-center extinction coefficients versus spatial depth in two model clouds with ionisation parameter, $\log U=-0.75$ but very different densities, $\log n_{\rm H} ({\rm cm}^{-3}) = 6$ and 10. Table 3 lists line-center optical depths and column densities four densities, $\log n_{\rm H} ({\rm cm}^{-3}) = 4$, 6, 8, and 10, all at $\log U=-0.75$. The locations of the three highest density cloud models in Table 3 are marked by magenta stars in Figure 7. 

\begin{figure*}
\includegraphics[scale=0.49]{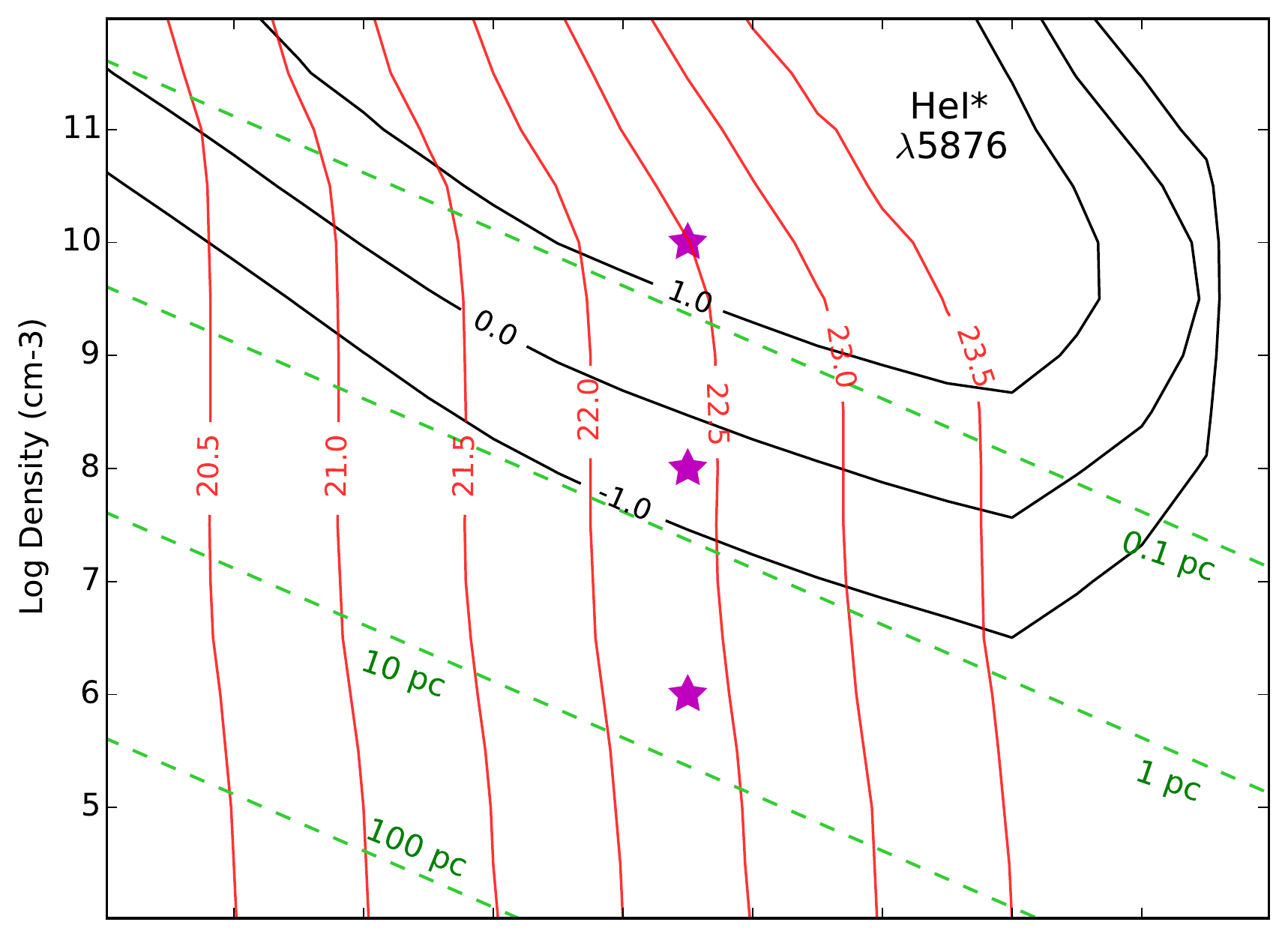}
\includegraphics[scale=0.49]{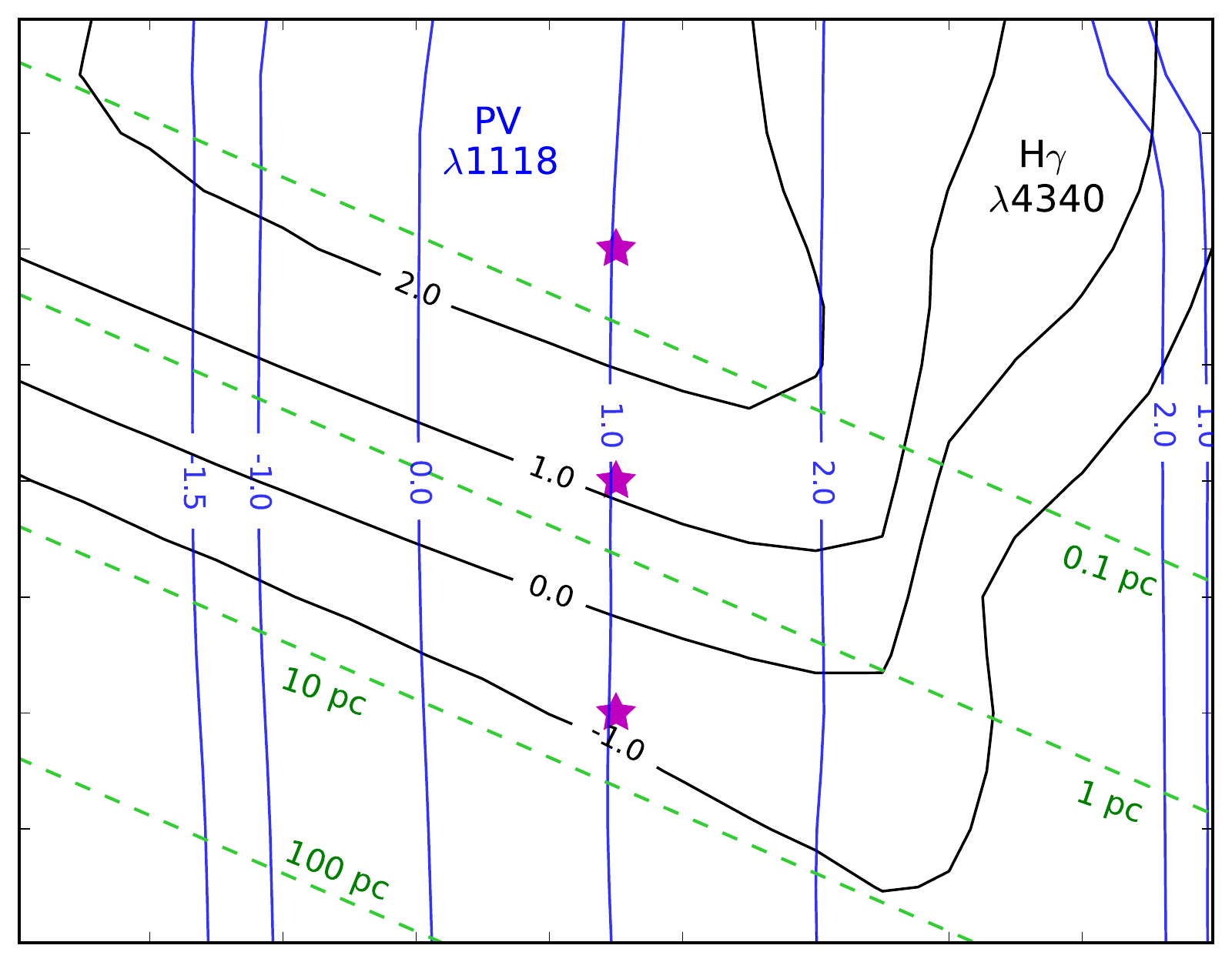}
\includegraphics[scale=0.49]{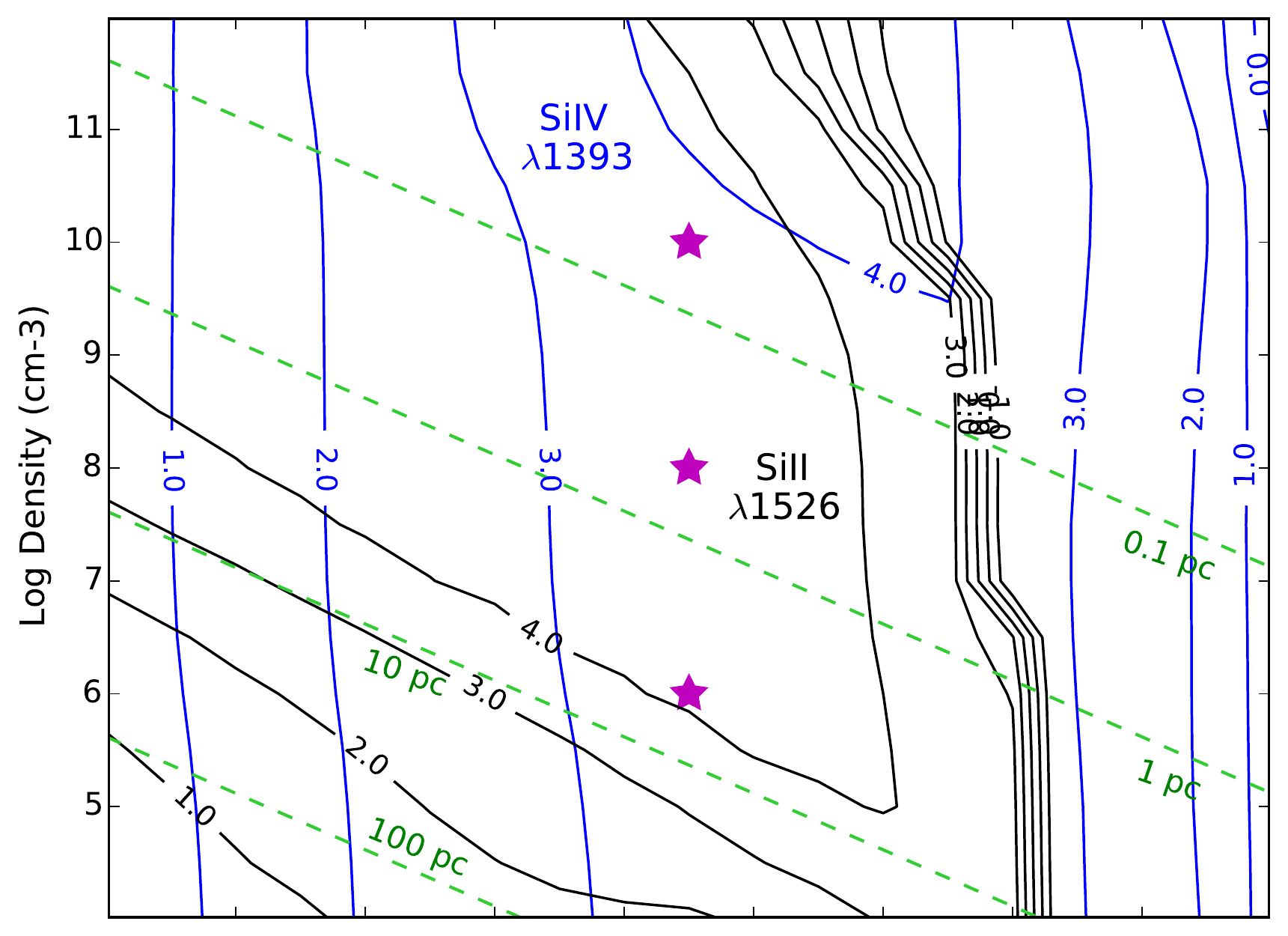}
\includegraphics[scale=0.49]{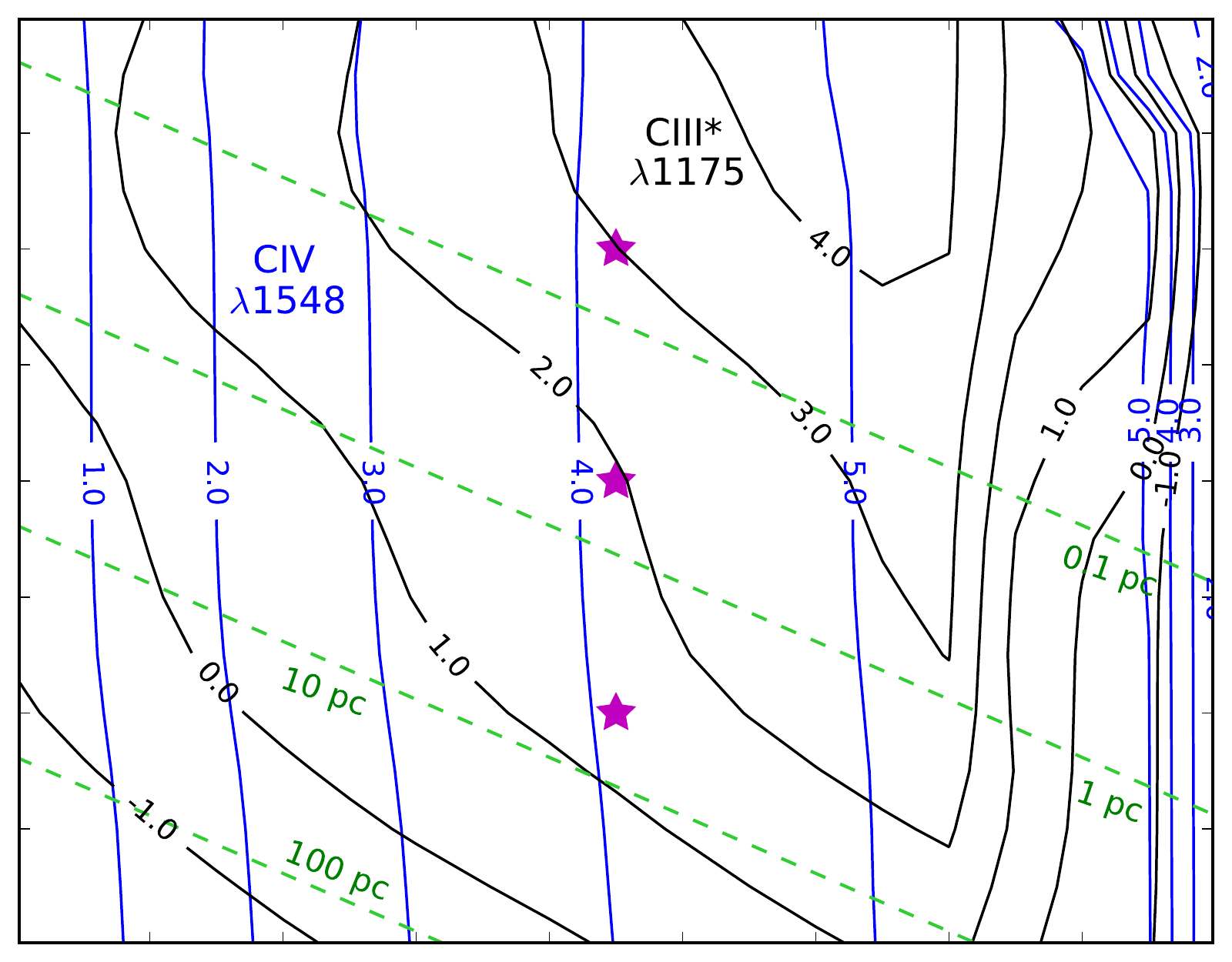}
\includegraphics[scale=0.49]{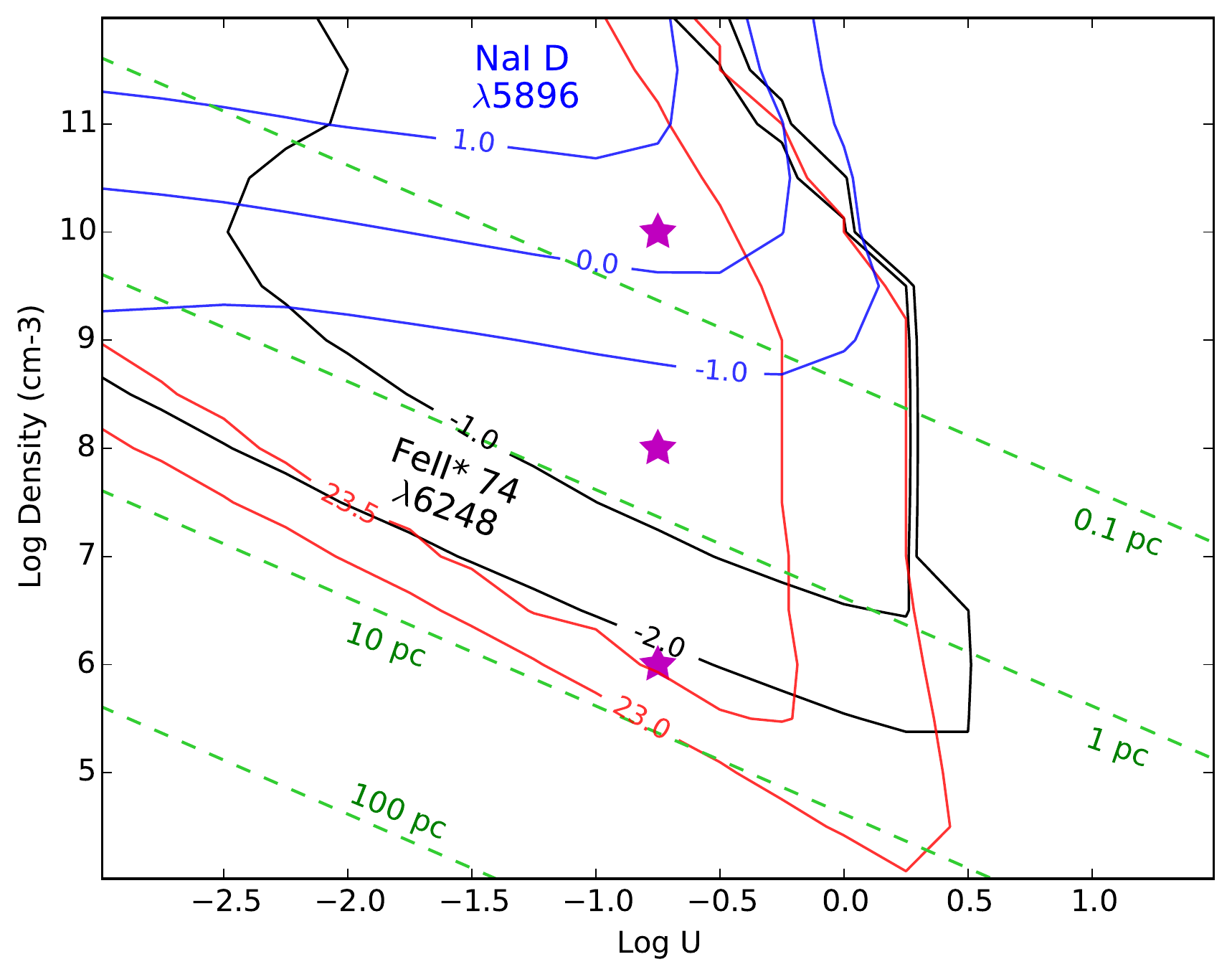}
\includegraphics[scale=0.49]{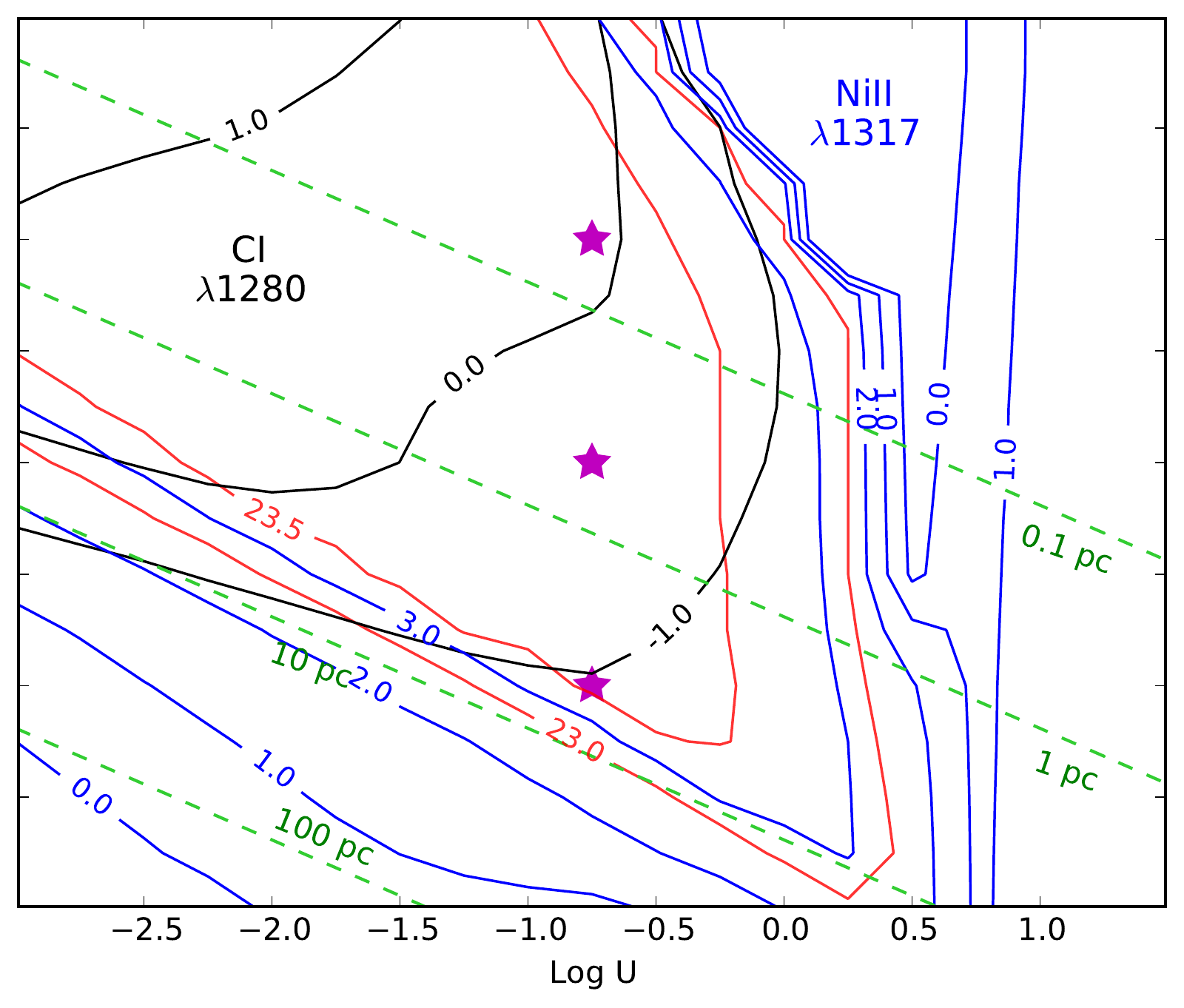}
\vspace{-8pt}
\caption{Theoretical line-center optical depths for clouds with $b=100$ \kms\ but different densities and ionisation parameters in our \cloudy\ simulations. Black and blue contours show log optical depths for the lines labeled in matching black or blue fonts. Red contours in the upper left panel indicate the hydrogen column density, $\log N_{\rm H} ({\rm cm}^{-2})$, in ionised gas before reaching the \heii --\hei\ recombination front; red contours in the bottom panels indicate the hydrogen column density in neutral and partially ionised gas behind this front. In the notation of Table 3, these column densities are $\log N_{\rm H}$(ionised) and $\log N_{\rm H}$(partial), respectively. Green dashed lines mark the distances of the clouds from the ionising continuum source (Equation 5). Magenta stars mark the models shown in Figure 8 and Table 3.}
\end{figure*}

\begin{figure*}
	\includegraphics[scale=0.51]{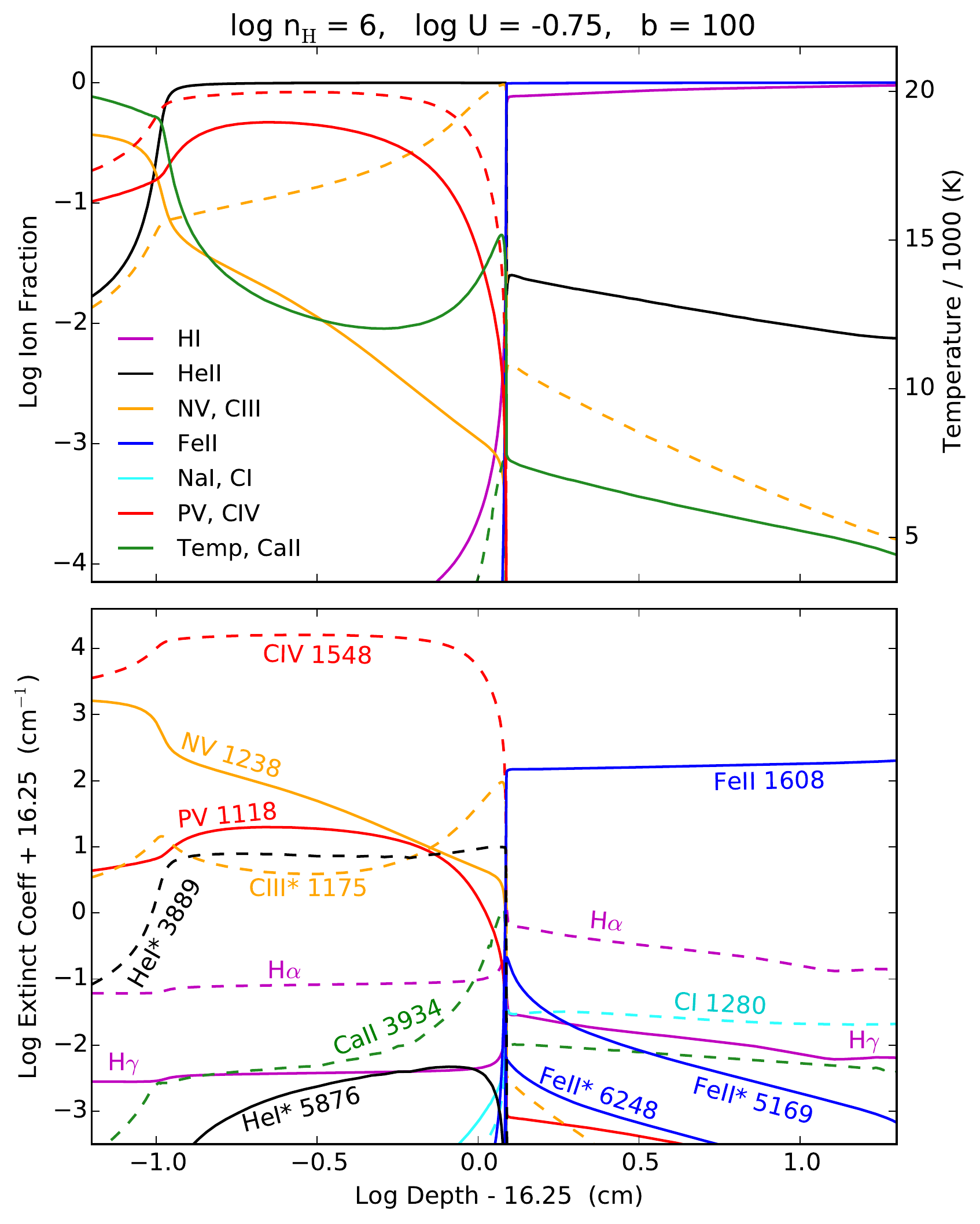}
	\includegraphics[scale=0.51]{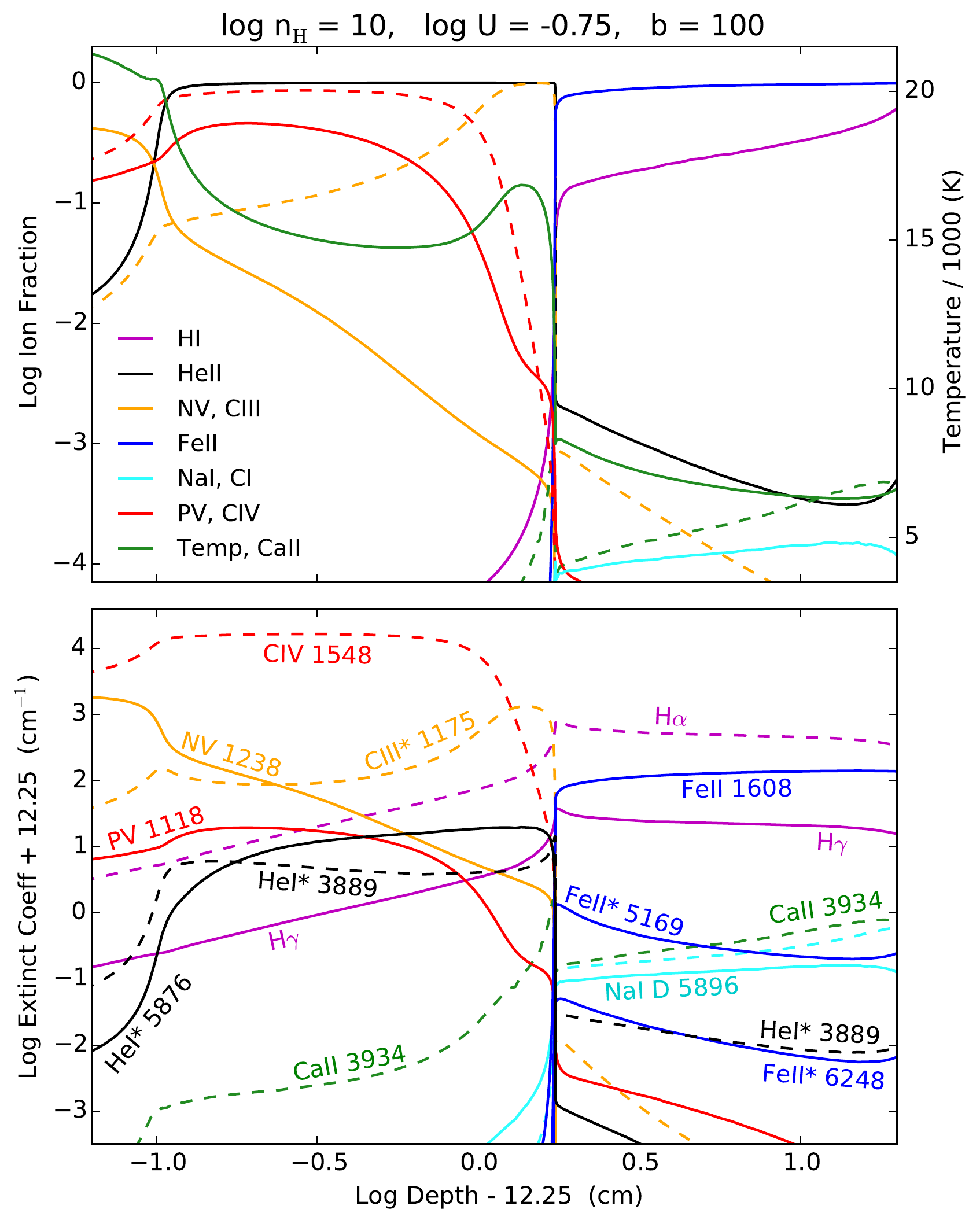}
\vspace{-8pt}
\caption{Ionisation fractions and line-center extinction coefficients versus spatial depth in two of the model clouds marked by magenta stars in Figure 6, both with $\log U = -0.75$ and $b=100$ \kms\ but different densities: $\log n_{\rm H} ({\rm cm}^{-3}) = 6$ (left panels) and 10 (right panels). The spatial depths are normalized so that 0.0 corresponds to the nominal location of the \hii --\hi\ recombination front at low densities: $\log N_{\rm H} ({\rm cm}^{-2}) \approx 23 + \log U = 22.25$. {\it Upper Panels}: Electron temperature (solid green curve) and various ionisation fractions indicated in the legend (if two ions have the same color, the first one listed is drawn with a solid curve while the second is dashed). {\it Bottom Panels}: Line-center extinction coefficients using the same color scheme as the top panel. The cloud with $\log n_{\rm H} ({\rm cm}^{-3}) = 10$ has higher temperatures and a deeper ionised zone due to enhanced excited-state photoionisations.  It also produces stronger absorption in excited-state lines of \hi, \hei* 5876, and \feii* as well as resonant lines of lower ions like \ci , \nai , and \caii . See also Table 3.}
\end{figure*}

\begin{table}
\centering
	\caption{Line-center optical depths for model clouds with $\log U = -0.75$, $b=100$ \kms , and $\log n_{\rm H} ({\rm cm}^{-3}) = 4$, 6, 8, and 10. The bottom rows give $\tau$(BaC) = continuum optical depth at the Balmer edge, $\log N_{\rm H}$(total) = total H column density (in \cmN ), and $\log N_{\rm H}$(ionised) = H column density in fully-ionised (H$^+$) gas. Higher densities produce much stronger absorption in excited-state lines of \hi*, \hei* 5876, and \feii* and resonance lines of neutrals like \ci , \mgi , and \nai . }
\tabcolsep=0.2cm
	\begin{tabular}{lcccc} 
\hline
%Line ID  & $\log n_{\rm H} = 6$& $\log n_{\rm H} = 8$& $\log n_{\rm H} = 10$\cr
 & \multicolumn{4}{c}{------------------------ \ \ $\tau_o$ \ \ ------------------------} \cr
\multicolumn{2}{r}{Line ID~~~~~~~~~~~~~~~~~~~$\log n_{\rm H} = 4$} & 6& $8$& $10$\cr
\hline
\lya\ 1216	&	7.55E+06	&	2.52E+09	&	2.38E+09	&	1.11E+09	\cr
\ha\ 6563	&	5.57E$-$02	&	4.07E+00	&	3.08E+02	&	9.10E+03	\cr
\hb\ 4861	&	7.66E$-$03	&	5.60E$-$01	&	4.24E+01	&	1.25E+03	\cr
\hg\ 4340	&	2.57E$-$03	&	1.88E$-$01	&	1.42E+01	&	4.20E+02	\cr
\hd\ 4102	&	1.20E$-$03	&	8.78E$-$02	&	6.65E+00	&	1.96E+02\vspace{2.5pt}	\cr
\hei* 3889	&	9.88E+00	&	9.03E+00	&	8.25E+00	&	8.21E+00	\cr
\hei* 10830	&	2.30E+02	&	2.10E+02	&	1.92E+02	&	1.91E+02	\cr
\hei* 5876	&	3.30E$-$05	&	3.52E$-$03	&	3.38E$-$01	&	2.60E+01\vspace{2.5pt}	\cr
\ci\ 1277	&	1.40E$-$03	&	4.58E$-$01	&	2.16E+00	&	7.73E+00	\cr
\ci\ 1280	&	2.78E$-$04	&	9.10E$-$02	&	4.30E$-$01	&	1.53E+00	\cr
\cii\ 1335	&	6.94E+02	&	2.30E+05	&	2.36E+05	&	2.32E+05	\cr
\ciii\ 977	&	1.85E+04	&	2.28E+04	&	2.55E+04	&	4.32E+04	\cr
\ciii * 1175	&	1.20E+00	&	2.64E+01	&	8.84E+01	&	9.63E+02	\cr
\civ\ 1548	&	1.03E+04	&	1.26E+04	&	1.40E+04	&	1.43E+04	\cr
\nii\ 1084	&	1.73E+02	&	4.19E+04	&	4.23E+04	&	5.40E+04	\cr
\nv\ 1238	&	1.69E+02	&	1.91E+02	&	2.11E+02	&	2.17E+02	\cr
\oi\ 1304	&	4.97E+02	&	1.68E+05	&	1.60E+05	&	8.09E+04	\cr
\ovi\ 1032	&	1.09E+02	&	1.33E+02	&	1.51E+02	&	1.46E+02	\cr
\mgi\ 2853	&	5.17E$-$01	&	4.33E$-$01	&	1.92E+00	&	8.55E+01	\cr
\mgii\ 2795	&	7.02E+02	&	3.09E+05	&	3.18E+05	&	3.01E+05	\cr
\nai~D 5896	&	4.34E$-$04	&	5.42E$-$04	&	1.45E$-$02	&	2.91E+00	\cr
\siii\ 1260	&	5.25E+02	&	1.03E+05	&	1.03E+05	&	1.01E+05	\cr
\siii * 1265 	&	4.25E+02	&	1.88E+05	&	1.95E+05	&	1.90E+05	\cr
\siii\ 1527	&	6.94E+01	&	1.36E+04	&	1.37E+04	&	1.34E+04	\cr
\siii * 1533 	&	5.62E+01	&	2.48E+04	&	2.58E+04	&	2.51E+04	\cr
\siiv\ 1394	&	1.68E+03	&	1.97E+03	&	2.26E+03	&	3.89E+03	\cr
\pv\ 1118	&	1.06E+01	&	1.09E+01	&	1.06E+01	&	1.05E+01	\cr
\sii\ 1254	&	2.76E+00	&	1.03E+03	&	1.06E+03	&	1.06E+03	\cr
\siv\ 1063	&	1.01E+02	&	6.19E+01	&	6.61E+01	&	8.95E+01	\cr
\siv * 1073	&	3.33E+01	&	9.85E+01	&	1.11E+02	&	1.51E+02	\cr
\caii\ 3934	&	3.22E$-$01	&	2.51E$-$01	&	2.89E$-$01	&	1.19E+01\vspace{2.5pt}	\cr
\feii\ ~ $a\,^6$D 1082	&	6.69E+00	&	8.30E+02	&	6.44E+02	&	6.20E+02	\cr
\feii* $a\,^6$D 1088	&	3.19E$-$01	&	3.61E+02	&	3.37E+02	&	3.24E+02	\cr
\feii* $a\,^6$D 1093	&	1.27E$-$01	&	1.92E+02	&	1.90E+02	&	1.83E+02	\cr
\feii\ UV8 1608	&	3.00E+01	&	3.72E+03	&	2.89E+03	&	2.78E+03	\cr
\feii\ UV1 2599	&	1.97E+02	&	2.44E+04	&	1.90E+04	&	1.83E+04	\cr
\feii*UV62 2749	&	3.58E$-$01	&	1.26E+03	&	3.01E+03	&	2.74E+03	\cr
\feii* 42 5169	&	5.59E$-$05	&	8.63E$-$02	&	4.52E+00	&	5.61E+00	\cr
\feii* 74 6248	&	9.19E$-$07	&	5.70E$-$03	&	2.68E$-$01	&	1.90E$-$01	\cr
\nkii\ $a\,^2$D 1317	&	5.78E+00	&	1.89E+03	&	1.94E+03	&	1.90E+03	\cr
\nkii\ $a\,^2$D 1346	&	2.59E$-$01	&	8.47E+01	&	8.70E+01	&	8.55E+01	\cr
\nkii\ $a\,^2$D 1502	&	5.88E$-$01	&	1.92E+02	&	1.97E+02	&	1.94E+02\vspace{2.5pt}	\cr
$\tau$(BaC)&  1.40e--05 &	9.24e--04  & 	6.98e--02 	& 	2.03e+00\vspace{2.5pt}\cr
$\log N_{\rm H}$(total)	&	22.28	&	23.59	&	23.60	&	23.60	\cr
$\log N_{\rm H}$(ionised)	&	22.26	&	22.34	&	22.38	&	22.49	\cr
\hline
\end{tabular}
\end{table}

The line extinction coefficients in Figure 8 ($l_{\textrm{v}}$, defined by $\tau_{\textrm{v}}~=~\int l_{\textrm{v}}\, ds$ where $s$ is the spatial depth) show where the lines form inside the model clouds. Clouds photoionised by the quasar spectrum adopted here have roughly co-spatial \hii --\hi\ and \heii --\hei\ recombination fronts at column density $\log N_{\rm H} ({\rm cm}^{-2})\approx 23 + \log U$ \citep{Davidson79}. The spatial depth scales in Figure 8 are normalized to this front location for easier comparisons between models. The actual front locations are evident from the abrupt decline in the \heii\ fraction accompanied by the steep rise in \hi\ and \feii . This occurs deeper in the clouds at higher densities because dense clouds have larger excited-state populations and, therefore, more ionisations out of excited states. We can see how high densities lead to enhanced excited-state photoionisations from the Balmer edge optical depths listed in Table 3, which grow from $\tau ({\rm BaC}) \sim 10^{-5}$ at $\log n_{\rm H} ({\rm cm}^{-3}) = 4$ to $\tau ({\rm BaC}) \sim 2$ at $\log n_{\rm H} ({\rm cm}^{-3}) = 10$. High densities also produce a much thicker zone of partially-ionised gas behind the \hii --\hi\ recombination front.

\subsection{Physical Conditions in the Transient Absorber}

Here we combine our \cloudy\ calculations with other analyses of the line strengths to estimate the densities, column densities, and ionisation structure in transient outflow. We focus mainly on well-measured lines in the visible because they are sufficient to place strong constraints on the outflow physical conditions, corroborated by the rich spectrum of transient lines in the UV.  

\subsubsection{Fully-ionised Gas: \hei* \lam 5876}

The \hei* \lam 5876 absorption line detected in the transient system requires both high densities and high column densities in fully-ionised gas. This line forms with other \hei* lines like \lam 3889 and \lam 10830 via recombination in regions dominated by \heii . The \lam 3889 and \lam10830 lines are not within our spectral coverage, but they should be stronger than \lam 5876 and they have been observed previously in some BAL and mini-BAL quasars \citep[e.g.,][]{Rupke02, Leighly11, Ji15}. \hei* \lam 3889 and \lam 10830 arise from the metastable $2s\,^3$S state at 19.8 eV. Significant detections in quasar spectra require large column densities (in spite of the large helium abundance) because the population in this highly-excited $2s\,^3$S state is always small compared to the ground state \citep[see][]{Leighly11}. \hei* \lam 5876 additionally requires large volume densities (frequent collisions) because it arises from a higher energy state that is {\it not} metastable, e.g., $2p\,^3$P$^o$ at 21.0 eV. This state is readily depopulated by permitted radiative decays to $2s\,^3$S via the \lam 10830 transition. Figure 8 and Table 3 show that the \hei* \lam 5876 optical depths (and the \lam 5876/\lam 3889 optical depth ratios) increase by nearly 4 orders of magnitude from $\log n_{\rm H} ({\rm cm}^{-3}) = 6$ to 10. The upper left panel in Figure 7 shows the explicit dependence of \lam 5876 on $n_{\rm H}$, $U$, and $N_{\rm H}$ in fully-ionised gas (red contours). We see from this plot that \lam 5876 line strengths corresponding to $\tau_o(\lambda 5876) \gtrsim 0.1$ require gas densities $\log n_{\rm H} ({\rm cm}^{-3}) \gtrsim 7$ for ionised column densities of $\log N_{\rm H} ({\rm cm}^{-2}) \sim 23$. More generally, this \lam 5876 optical depth in the transient absorber (with $b=100$ \kms) requires $\log n_{\rm H} ({\rm cm}^{-3}) \gtrsim (7+\log U)$ for $\log N_{\rm H} ({\rm cm}^{-2}) \sim (23+\log U)$ and any $U$ in the range $-2.0 \lesssim \log U \lesssim +0.5$. 

The large column densities needed for \hei* \lam 5876 absorption imply that all of the more common resonance lines of higher ionization ions are extremely optically thick. Figures 7 and 8 show that the \hei* \lam 5876 line-forming region overlaps with that of the resonance lines \civ\ \lam 1548, \siiv\ \lam 1393, and \pv\ \lam 1118. The relationship of the \hei* \lam 5876 line strength to these resonance lines is not straightforward because \lam 5876 has an additional density dependence not seen in the resonance lines. However, for all of the model clouds with $\log U = -0.75$ in Table 3, the densest of which are plausibly like the transient absorber in PG~1411+442, the line-center optical depths in \civ , \siiv , and \pv\ are $\sim$14,000, $\sim$3000, and $\sim$11, respectively.  These highly saturated resonance lines in the transient absorber are not clearly evident in the observed 2015 spectrum (Figures 2 and 3) because of 1) the small covering fraction of the transient absorber, and 2) blending with the broader and deeper mini-BALs produced by these same ions in the larger mini-BAL outflow (Section 5.6). 

\subsubsection{Partially-ionised Gas: \hi* and \feii* lines}

The \feii* and \hi* Balmer absorption lines provide strong constraints on the densities and column densities in partially-ionised gas behind the \hii --\hi\ front (Figure 8). First note that the apparent optical depths listed for the Balmer lines in Table 2 are very different from the predicted ratio of 1.0\,:\,3.00\,:\,21.7 for \hg\,:\,\hb\,:\,\ha . Saturation in these lines should push the observed ratio toward unity, but the data yield an unphysical result for $\tau_o$(\ha ) $< \tau_o$(\hg ). We attribute this to the combined effects of line saturation {\it and} dilution of the absorption-line strengths by unabsorbed flux from the underlying broad Balmer emission lines \citep{Ganguly99}. The amount of dilution is larger for the \ha\ absorption line because it sits on the stronger \ha\ emission line (Figures 4 and 5). We correct for this dilution by subtracting off the unabsorbed broad emission-line flux at the absorption-line wavelengths. The red curves in Figure 6 show the corrected Balmer line fits, while $\tau_o^c$ in the Notes column in Table 2 gives the corresponding corrected apparent optical depths at line center. These corrections are approximate because they do not account for other effects such as the wavelength-dependent size of the continuum source \citep{Shi17}, line-dependent size of the Balmer broad emission-line region, and the absorber geometry in relation to these emitting regions. None the less, the simple corrections shown in Figure 6 clearly account for the main line ratio anomaly. The revised line strength ratio for \hg\,:\,\ha\ is now $\sim$1.0\,:\,1.3, which clearly indicates that at least the \ha\ absorption line is highly saturated. The specific optical depths needed to produce this ratio are $\tau_o$(\hg )~$\gtrsim 1.3$ and $\tau_o$(\ha )~$\gtrsim 28$ if both lines have the same covering fraction. These are lower limits because \hg\ could also be optically thick while remaining weaker than \ha\ due to optical depth-dependent covering fractions in an inhomogeneous absorber \citep[see Section 5.5, also][]{Hamann04, Arav05, Arav08}. 

Thus the Balmer absorption lines are highly saturated. The top right panel in Figure 7 shows how the requirement for $\tau_o$(\hg )~$\gtrsim 1.3$ constrains the gas densities. The bottom panels show (in red contours) the column densities of partially-ionised gas that produce this absorption. The specific density constraints depend on the ionisation parameter, but for any $U$ value $\tau_o$(\hg )~$\gtrsim 1.3$ requires $\log n_{\rm H} ({\rm cm}^{-3}) \gtrsim 6.5$ and $\log N_{\rm H} ({\rm cm}^{-2}) \gtrsim 23.2$ in partially-ionised gas. 

The \feii* lines require more extreme conditions than \hg. The visible \feii* lines (Table 2) arise from highly-excited metastable states populated by collisions in a warm dense partially-ionised gas. High densities are needed to maintain the populations in these levels and large column densities are needed to produce measurable \feii* absorption. The bottom left panel in Figure 7 shows the values of $n_{\rm H}$ and $N_{\rm H}$ needed for significant optical depths in \feii* 74 \lam 6248 whose lower state has energy 3.89 eV. Results for other \feii\ and \feii* lines are shown in Table 3 and Figure 8. 

These \cloudy\ simulations also show that the Fe$^+$ level populations up to 3.89 eV are within factors of 2 to 3 of thermal equilibrium for densities $\log n_{\rm H} ({\rm cm}^{-3}) \gtrsim 7$ and within $\sim$20 percent for $\log n_{\rm H} ({\rm cm}^{-3}) \gtrsim 8$. Figure 9 plots predicted line-center optical depths for \feii* 42 \lam 5169, \feii* 74 \lam 6248, and the resonance line \feii\ UV8 \lam 1608 for idealised clouds in strict thermal equilibrium and with favorable ionisation conditions where all of the iron is Fe$^+$. The level populations are determined from the Boltzmann equation at temperatures $T = 5500$, 7000, and 8500 K. These simple calculations illustrate how the \feii* line strengths depend on $N_{\rm H}$ in dense environments with other parameters fixed. If the actual densities are not high enough to achieve thermal equilibrium, then excited-state populations would be smaller and even larger column densities than those shown in Figure 9 would be needed to produce these excited-state absorption lines. 

\begin{figure}
	\includegraphics[scale=0.48]{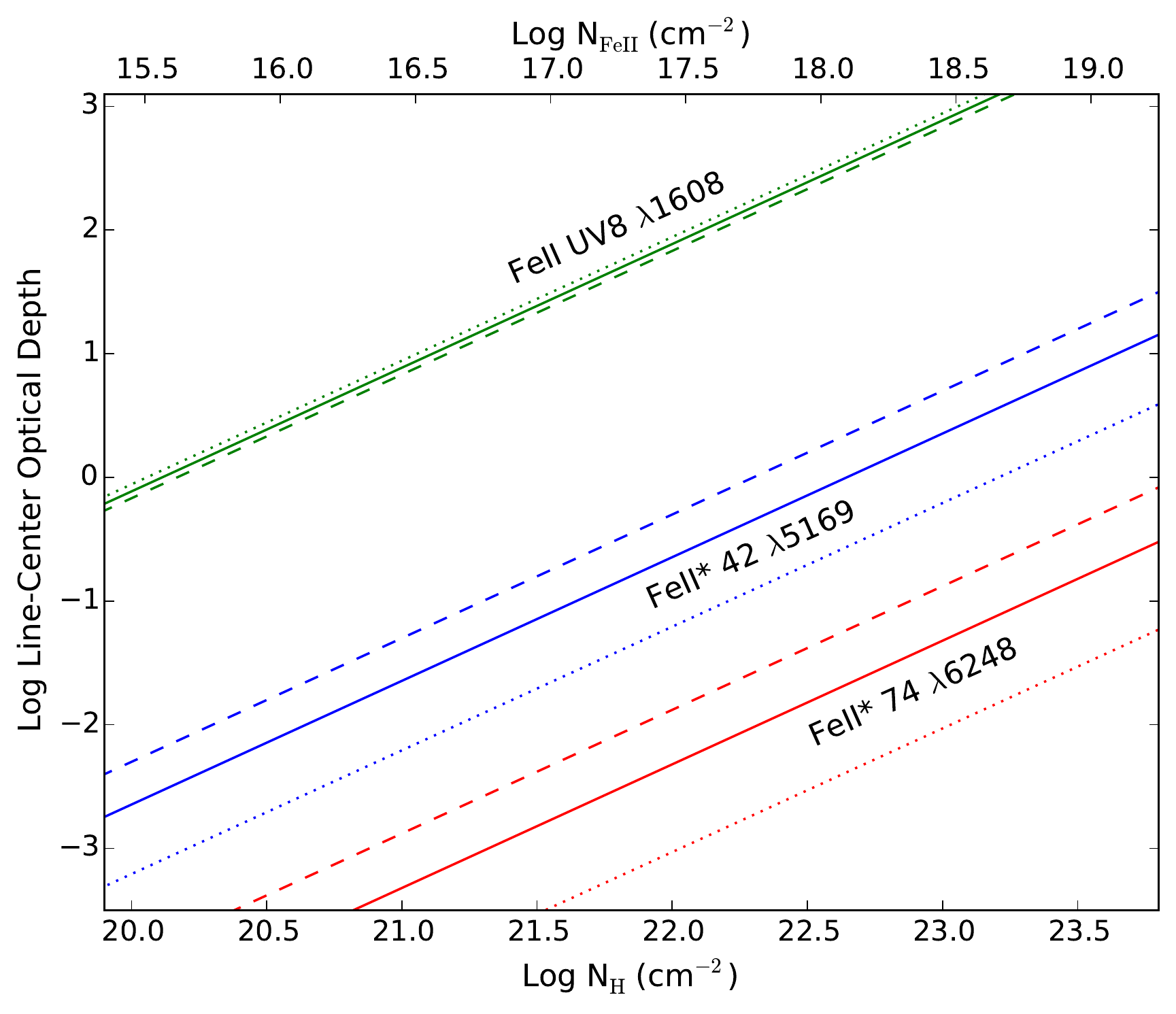}
\vspace{-12pt}
\caption{Theoretical line-center optical depths in three \feii\ lines as a function of total column density, $N_{\rm H}$, and \feii\ column density, $N_{\rm FeII}$, in clouds with solar abundances, Doppler parameter $b=100$ \kms , level populations in thermal equilibrium, and ionisation fraction Fe$^+$/Fe = 1. The three curves for each absorption line (dotted, solid, dashed) indicate temperatures $T = 5500$, 7000, and 8500 K (respectively). Note that the optical depth $\tau(\lambda 6248) \gtrsim 0.2$ we infer for the transient absorber requires $\log N_{\rm H} ({\rm cm}^{-2}) \gtrsim 23.3$ for any $T\lesssim 8500$ K.
}
\end{figure}

We can see from Figure 9 that specific constraints on the total column densities depend on the line optical depths. The \cloudy\ results in Table 3 and our idealized calculations in Figure 9 both indicate that the optical depths in \feii* 42 \lam 5169 and \feii\ UV8 \lam 1608 are at least $\sim$35 and $\sim$$1.5\times 10^4$ times larger than \feii* 74 \lam 6248, respectively. This contrasts dramatically with the measured apparent optical depth ratios of $\tau_o(\lambda 5169)/\tau_o(\lambda 6248)\sim 5.4$ and $\tau_o(\lambda 1608)/\tau_o(\lambda 6248)\sim 34$, respectively (Table 2), confirming that the lower-state lines are very optically thick. Following our analysis of the Balmer lines above, we find that the specific line optical depths needed to match the observed ratios are $\tau_o(\lambda 6248) \gtrsim 0.2$, $\tau_o(\lambda 5169) \gtrsim 7$, and $\tau_o(\lambda 1608) \gtrsim 3000$. Figures 9 and 7 (bottom left panel) then show that $\tau_o(\lambda 6248) \gtrsim 0.2$ requires total column densities $\log N_{\rm H} ({\rm cm}^{-2}) \gtrsim 23.3$ and densities $\log n_{\rm H} ({\rm cm}^{-3}) \gtrsim 7$ in the partially-ionised Fe$^+$ zone. 

It is important to note that the creation of this high column density partially-ionised Fe$^+$ zone requires an intense radiation field (with large $U$) shielded behind an \hii --\hi\ recombination front, i.e., behind a thick layer of fully-ionised gas. In particular, Figure 7 shows that $\tau_o(\lambda 6248) \gtrsim 0.2$ requires conservatively $\log U \gtrsim -2$ and $\log N_{\rm H} ({\rm cm}^{-2}) \gtrsim 21$ in fully-ionised gas. This shielding gas is very likely the \hei* absorber described above in Section 5.4.1, such that the minimum total column density through the entire transient absorber (ionised and partially-ionised) is $\log N_{\rm H} ({\rm cm}^{-2}) \gtrsim 23.4$ \cmN . 

\subsubsection{Neutral Atoms: \nai~D lines}

The \nai~D lines in the transient absorber pose an interesting problem because the \nai\ ionisation energy (5.1 eV at 2432 \AA ) is too low to be radiatively shielded behind a \hi --\hii\ recombination front that blocks photons above 13.6 eV. A similar problem exists for \ci\ (tentatively detected in \lam 1277 and \lam 1280) with ionisation energy 11.3 eV. 

How do these neutral atoms survive in the intense quasar radiation field? Two possibilities are strongly disfavored for the outflow in PG~1411+442. One is that the \nai~D lines form in a separate neutral environment with very low $U$, perhaps farther from the quasar than the other transient lines. This is unlikely given the similar strengths, velocity profiles, and temporal behaviors of the \nai~D lines compared to other lines in the transient absorber (see also HTRV2). Another possibility is that dust is present to block ionising radiation at wavelengths $\lesssim$2432 \AA . If this dust fully covered the continuum source along our lines sight, the quasar should appear red and faint in the UV. This is not seen. If dust is mixed with the absorption-line gas, the transient lines should be weaker at shorter wavelengths, depending on the shape of the reddening curve \citep{Veilleux13a}. In fact, the transient lines show the opposite trend, where optically-thick lines in the UV are stronger than in the visible (Table 2, Figures 2 to 5). Dust inside the absorber could also cause problems by depleting the iron abundance and inhibiting the formation of the thick partially-ionised zone needed to form the observed \feii\ and \feii* transient lines. 

A more likely explanation is that \nai~D lines form in the same partially-ionised gas as the \feii* lines, but the densities in that region are higher than the lower limits we derived in Section 5.4.2. The \cloudy\ results in Table 3 and Figure 7 illustrate how the \nai~D optical depths increase by almost 4 orders of magnitude from $\log n_{\rm H} ({\rm cm}^{-3}) = 6$ to 10. This is due partly to increasing radiative shielding in the near-UV caused by excited-state bound-free opacities in a variety of ions. Notice, in particular, the dramatic rise in the \hi\ Balmer-edge optical depths (indicating larger $n=2$ populations) with increasing density in Table 3. Another factor is that higher electron densities in the partially-ionised zone increase the rate of \naii\ to \nai\ recombinations to produce more \nai\ and, therefore, stronger \nai~D lines. 

The \nai~D lines have an intrinsic optical depth ratio of $\tau (\lambda 5890)/\tau (\lambda 5896) \approx 2$. The observed ratio (Table 2) is indistinguishable from unity if we consider that, like the Balmer lines, the \lam 5896 line strength is diluted more than \lam 5890 by underlying broad \hei\ \lam 5876 emission (Figure 4). Thus the observed ratio implies that the \nai~D lines are saturated, with $\tau_o(\lambda 5896) \gtrsim 3$. The results in Figure 7 and Table 3 then show that very high densities, $\log n_{\rm H} ({\rm cm}^{-3}) \gtrsim 10$, are required.

\subsubsection{\siii\ and \siii * lines}

\siii\ and \siii * lines pairs can be valuable density diagnostics in absorption-line environments. We attribute the \siii\ and \siii * lines observed in 2015 primarily to the mini-BAL outflow because 1) they were present in 2011 before the transient lines appeared, and 2) the \siii\ \lam 1260 and \siii * \lam 1265 lines broadened in 2015 like other mini-BALs, such that their widths, along with the \siii\ \lam 1527 and \siii * \lam 1533 lines, were roughly twice as broad in 2015 as the \feii , \feii*, and \nkii\ in the transient absorber (Section 4, Table 2). Nonetheless, we examine the \siii\ and \siii * lines to constrain the densities near the ambiguous boundary between the transient absorber and the mini-BAL outflow.

The \siii * lines arise from a fine structure level 0.036 eV above the ground state. The critical density of this level is $\log n_{cr} ({\rm cm}^{-3}) \sim 3.1$ for a nominal temperature of $T\approx 7000$K \citep[using atomic data from NIST and][]{Tayal08}. We calculate \siii */\siii\ optical depth ratios as a function of density assuming only collisional processes and radiative decays in a 2-level atom \citep[see also][]{Moe09, Dunn10}. The predicted ratios for \lam 1265/\lam 1260 and \lam 1533/\lam 1527 range from $\sim$0 well below the critical density to $\sim$2 at $\log n_{H} ({\rm cm}^{-3}) \gtrsim 4.2$. (The \siii\ and \siii * lines near $\sim$1193 \AA\ are not useful in our spectra because of blending problems.) If the lines are optically thick, the measured line ratios will approach unity from either of these limits. 

In our 2011 spectrum, the resonant \lam 1260 line is poorly measured but the \lam 1265/\lam 1260 strength ratio is clearly large, e.g., in the high-density regime with $\log n_{H} ({\rm cm}^{-3}) \gtrsim 4.2$. This is consistent with the minimum densities $\log n_e ({\rm cm}^{-3}) \gtrsim 5$ we derive from the \siv * and \ciii * mini-BALs (HTRV2). 

In 2015, the \lam 1265/1260 line ratio appeared crudely close to unity but it is also mired in blends. Also in 2015, the \siii /\siii * \lam 1533/\lam 1527 ratio appears well-measured at a value of $\sim$$0.6\pm 0.1$ (Figure 2, Table 2). This ratio is surprising because it seems to indicate relatively low densities, $\log n_e ({\rm cm}^{-3}) \lesssim 3.4$, regardless of saturation effects. This much lower density compared to our estimates for the transient absorber above might be another indication of the complex highly-structured nature of the PG~1411+442 outflow (Section 6). {\it However}, there are three \feii* lines blended with \siii\ \lam 1527 that should contribute significantly to its measured strength. The \feii* lines are 1526.54 \AA\ with strength index $P=1.06$ and 1527.30 \AA\ and 1527.35 \AA\ with combined $P=0.89$ (see Figure 6 and our line list from Section 5.1). Other \feii* lines with similar $P$ values are clearly detected in the 2015 spectrum. Inspection of those lines suggests that altogether \feii* contributes $\sim$1/3 to the measured strength of \siii\ \lam 1527. A weaker \feii* line with $P=0.54$ blended with \siii * \lam 1533 should contribute negligibly to that line. Therefore, the true \siii /\siii * \lam 1533/\lam 1527 strength ratio in 2015 appears to be of order unity, consistent with saturation but not useful for density estimates. This conclusion is supported by subsequent HST-COS observations where the transient features are weaker (so there should be less contamination by \feii*) and the \textit{observed} \lam 1533/\lam 1527 ratio is roughly unity (HTRV2). 

\subsection{Radial Distance \& Revised Density Limits}

The dashed green curves in Figure 7 indicate the distance of the absorbing gas from the quasar continuum source for different values of $U$ and $n_{\rm H}$ (from Equation 5). They show that the distances must be small, conservatively $\lesssim$0.4 pc to produce the \feii* 74 and \nai~D lines and $\lesssim$1.0 pc for \hei* \lam 5876. 

A more strict location constraint comes from our conclusion in Section 5.4.2 that the \hi* transient lines do not cover the \hi\ broad emission-line region. This places the absorber nominally within the BLR radius at $R_{BLR} \approx 0.10\pm 0.05$ pc (Section 3, although larger radii might be allowed for particular geometries and viewing perspectives). If we adopt this smaller distance (corresponding to $\sim$6000 gravitational radii), then the densities required by the \hi* and \feii* lines are larger, e.g., the \hg\ optical depth of $\tau _o \gtrsim 1.3$ now requires $\log n_{\rm H} ({\rm cm}^{-3}) \gtrsim 8.2$ while $\tau_o \gtrsim 0.2$ in \feii* 74 \lam 6248 requires $\log n_{\rm H} ({\rm cm}^{-3}) \gtrsim 8.7$ (Section 5.4.2 and Figure 7). The density lower limit from the \nai~D lines remains unchanged at $\log n_{\rm H} ({\rm cm}^{-3}) \gtrsim 10$ (Section 5.4.3). 

\subsection{Partial Covering \& Spatial Structure}

We have seen that most of the transient lines are optically thick in spite of their weak appearance in the spectrum. We attribute this to partial covering of the quasar continuum source. In principle, scattered light could also fill in the absorption-line troughs to produce similar effects \citep[e.g.,][]{Kraemer01}. However, the deep \civ\ mini-BAL in PG~1411+442 indicates that the scattered light contributions are $\lesssim$8 percent of the continuum (HTRV2). This is much less than what is needed to explain the observed depths of optically-thick transient lines, which would require scattering contributions of $\sim$60 to $\sim$75 percent in the UV up to $\sim$96 percent in the visible (Table 2 and Section 5.4.2). Another problem with the scattered light explanation is that different optically-thick lines have different observed depths even at similar wavelengths. This seems wholly incompatible with the scattering hypothesis, but it can be explained naturally by inhomogeneous partial covering. 

Different covering fractions in different lines are, in fact, a signature of a spatially-inhomogeneous absorber \citep{deKool02, Arav05, Hamann01, Hamann19, Hamann04, Gabel05, Leighly11, Chamberlain15, Moravec17}. If the absorber spans a wide range of line optical depths from thick to thin across the projected area of the emission source, stronger lines of abundant ions will be optically thick over larger spatial areas, leading to larger covering fractions and stronger/deeper absorption troughs in observed spectra. Cartoon illustrations of this absorbing geometry can be found in \cite{Hamann01} and  \cite{Hamann04}. The observed depths of saturated lines are direct indicators of the covering fractions of optically-thick gas in that transition (Equation 3). 

In the transient absorber, the measured line depths range from $\sim$0.04 in the \nai~D lines (after a small adjustment for broad \hei\ \lam 5876 emission slightly filling in their troughs, see Figure 5 and Section 5.4.3), to $\sim$0.07 for \feii* 42, $\sim$0.19 for the \hi* Balmer lines, and $\sim$0.45 for \feii\ UV8 \lam 1608 and other lines in the UV (Figures 2 and 5). The weak optically-thick \nai~D lines in indicate that the extreme physical conditions needed for these lines (large densities and column densities, Section 5.4.3) occur only in small spatial regions, while strong resonance lines like \feii\ UV8 \lam 1608 can form in a wider range of environments that present larger covering fractions. 

A caveat to this discussion is that the accretion-disk emission source is smaller at shorter wavelengths. The predicted \hei* lines \lam 10830 and \lam 3889 could be valuable diagnostics of this effect because they span a wide range of wavelengths while arising from the same lower energy state \citep[see][]{Leighly11}. We do not cover those lines nor anything equivalent in the transient system. However, the generally larger covering fractions in the UV lines compared to the visible might be affected by the smaller size of the UV continuum source.

We convert the absorber covering fractions to physical dimensions by considering the size of the continuum emission source. Theoretical accretion disks that are optically-thick and geometrically-thin have emission region radii that scale with wavelength roughly as 
\begin{equation}
R_{c} \ \approx \ 10^{16}\, \left({f_E}
\over{0.4}\right)^{1/3}\left({{M_{BH}}\over{10^9\,{\rm M}_{\odot}}}\right)^{2/3}
\left({{0.1}\over{\eta}}\right)^{1/3}\left({{\lambda_{p}}\over{1500\,{\rm\AA}}}\right)^{4/3}~~{\rm cm}
\end{equation}
where $f_E = L/L_E$ is the Eddington ratio, $M_{BH}$ is the black hole mass, $\eta$ is the radiative efficiency factor, and $\lambda_{p}$ is the peak wavelength of the Planck function at the local disk temperature \citep{Peterson97}. The best measure of the emission-region sizes for a covering fraction analysis is probably the half-light radius. However, we use Equation 6 because it gives radii intermediate between theoretical half-light radii for optically-thick geometrically-thin accretion disks \citep[which are $\sim$3.7 times smaller than Equation 6,][]{Blackburne11} and observed half-light radii inferred from micro-lensing studies \citep[crudely $\sim$2 times larger than Equation 6, but with a wide range,][]{Chartas16,Blackburne15,Blackburne11,Morgan10}. Plugging in the black hole mass and Eddington ratio for PG~1411+442 (Section 3) yields continuum emission radii of $R_c\sim 1.8\times 10^{15}$ cm at 1100 \AA\ and $\sim$$1.8\times 10^{16}$ cm at 6200 \AA . 

If we imagine for a moment that the transient absorber is a circular disk projected onto a circular emission source with uniform brightness, then the UV covering fractions up to $C_{\textrm{v}}\sim 0.45$ correspond to an absorber radius of $r\sim 1.2\times 10^{15}$ cm in front of the UV light source, while $C_{\textrm{v}}\sim 0.04$ to 0.19 in the visible lines indicates $r\sim (4~{\rm to}~ 8)\times 10^{15}$ cm in front of the visible light source. These are only characteristic sizes, but they indicate that the visible light absorber is larger than the entire UV emission source. The actual absorber geometry needed to explain these results is certainly not a circular disk. It might be elongated radially, perhaps in outflow streams, and/or an aggregate of many clouds that are each smaller than the smallest size derived above. 

Another constraint on the spatial structure of the transient absorber comes from the derived densities and column densities. If we adopt a total column density of $\log N_{\rm H} ({\rm cm}^{-2}) \sim 23.5$ compatible with our analysis in Sections 5.4.1 and 5.4.2 and a density of $\log n_{\rm H} ({\rm cm}^{-3}) \sim 10$ from the \nai~D lines (Section 5.4.3), then the radial thickness of a contiguous absorber is only $\Delta r \sim N_{\rm H}/n_{\rm H} \sim 3\times 10^{13}$ cm. This is $\gtrsim$150 times smaller than the transverse diameter covering the visible light source, indicating that the transient absorber has the shape of a thin ``pancake'' flattened in the radial direction or it is spatially distributed in many small clouds with individual sizes $<$$3\times 10^{13}$ cm across. The latter possibility of outflows composed of many small clouds is consistent with some recent outflow studies \citep[][and refs. therein]{Hamann13, Baskin13, Baskin14} and the longstanding picture of quasar BLRs \citep[see \S6, also][]{Ferland92, Arav97, Dietrich99, Bottorff00}. 

\subsection{Kinematics \& Emergence of the Transient Absorber}

The transient absorption lines emerged on a time scale of $<$3.04 years in the quasar frame between our HST observations in 2011 and  2015  (Table 1, Figures 2 and 3). What caused these features to appear? The possibilities usually discussed for outflow line variabilities are ionisation changes caused by changes in the incident quasar flux, or transverse motions of outflow clouds across our lines of sight to the continuum source. Discussions of these two possibilities in the literature are often oversimplified. Transverse cloud motions do not mean simply changes in the absorber covering fractions at a fixed ionisation or fixed column density. All of those parameters can change in unpredictable ways if the moving absorber is clumpy and inhomogeneous, such that clouds with different ionisations and different column densities appear at different time. Another complication is that changes in the ionising flux can produce changes in the covering fractions without transverse motions, e.g., as the lines become optically thick over larger/smaller areas in a clumpy inhomogeneous absorber \citep[see Figure 4 in][]{Hamann12}. Thus the two different explanations for the line variability can produce similar observed results. 

Nonetheless, attributing the emergence of the transient absorber to changes in the incident flux is problematic for two reasons. First, the observed UV continuum flux changed by only $\sim$20 to $\sim$25 percent between 2011 and 2015. Also, the UV flux in 2011, when the transient lines were absent, was intermediate between two subsequent observations in 2015 and 2016 that both showed the transient lines (HTRV2). Another observation in 2017 shows the transient lines nearly disappearing while the continuum flux went back down near the 2015 levels (when the transient lines were strongest). Thus the continuum behavior shows no relationship to the emergence of the transient lines. 

%It might be possible to induce ionisation changes in the outflow without changes in the near-UV flux if there is an external shielding medium at the base of the outflow that blocks far-UV radiation via bound-free opacities \citep{MisawaXX, Wang15}. However, an external shield is not likely to play a significant role for the reasons discussed in Section 5.3 \citep[also][]{Hamann13}.

The second problem is that many of the transient lines are very saturated, with optical depths in the hundreds or thousands for UV lines like \nkii\ \lam 1317, \feii\ \lam 1082 and \lam 1608, and \feii* \lam 1088 and \lam 1093 (Table 3 and Figure 7). The only way to induce large changes in the line optical depths with small/moderate changes in the ionising flux is if the absorber had specific values of $U$ and $N_{\rm H}$ to place the \hi --\hii\ recombination front near the back of the clouds, so that smaller $U$ values cause the clouds to transition from ionisation-bounded (no partially-ionised zone) to matter-bounded (with a partially-ionsed zone). In our calculations, this threshold occurs near $\log U\sim 0.3$. However, this situation seems very unlikely because it requires fine tuning to produce the observed emergence of the transient lines. 

Another possible cause of absorption-line variability is the dissipation of small absorbing clouds/clumps in the outflow. If the clouds are not confined by an external pressure, they will dissipate in roughly a crossing time for particles moving at an internal cloud speed. The maximum internal velocity given by the doppler line widths, $b \sim 100$ \kms\ (Table 2) combined with the radial thickness, $\Delta r$, from Section 5.5 yields a dissipation time of $\Delta r/b \gtrsim 1$ month. This is easily small enough to affect the line strengths within our observation period. However, dissipation means lower densities and the eventual loss of the cloud, which is the opposite of what is needed to explain the \textit{emergence} of the low-ionisation transient absorber in 2015. 

We conclude that the transient lines appeared due to transverse motions in the outflow leading to small, dense, high-column density clouds entering our lines of sight to the continuum source. If we adopt a nominal transverse absorber diameter of $d \sim 6 \times 10^{15}$ cm  from Section 5.6, then the minimum crossing speed needed to move from fully-off to fully-on the continuum source in the time $\Delta t < 3.04$ years is $d/\Delta t \gtrsim 625$ \kms\ \citep[see also][]{Capellupo14}. This is less than the Keplerian rotation speeds, $\gtrsim$2180 \kms , at the estimated absorber location $\lesssim$0.4 pc from the black hole (Sections 3 and 5.5). Therefore, within the limited constraints of our time sampling, the emergence of the transient absorber is consistent with the crossing speeds expected for an outflow still close to the rotating accretion disk. 

These same calculations indicate that the measured outflow speed, |v|$\,\sim 1900$ \kms , is less than gravitational escape speeds, v$_{esc}\gtrsim 3080$ \kms , at the radial distance $\lesssim$0.4 pc from the black hole. Therefore, surprisingly, the transient ``outflow'' in PG~1411+442 is gravitationally bound to the quasar. 

\section{Summary \& Discussion}

We describe the emergence and physical properties of a unique narrow absorption-line outflow in the quasar PG~1411+442. This new ``transient'' outflow system contains an estimated $\geq$2990 absorption lines at rest wavelengths from 940 \AA\ to 6500 \AA , dominated by \feii\ and \feii* from excited states up to 3.89 eV. Includes includes many rare features like the \hi* Balmer lines and \nai~D \lam 5890,5896, plus \hei* \lam 5876 absorption and several visible \feii* multiplets that are measured here for the first time in a quasar outflow. The transient system appeared at roughly the same velocity as the mini-BAL outflow measured previously at v$\,\sim -1900$ \kms , but the transient lines are weaker, narrower at $b\sim 100$ \kms , and they require more extreme physical conditions with larger densities and larger column densities in a smaller outflow volume. Our analysis of the transient system yields the following results:

(1) The lines measured in the transient system require an intense radiation field near the quasar to generate large column densities in both partially-ionised and fully-ionised gas on either side of an \hii --\hi\ recombination front. The minimum column densities in these regions are $\log N_{\rm H} ({\rm cm}^{-2}) \gtrsim 23.3$ and $\log N_{\rm H} ({\rm cm}^{-2}) \sim (23+\log U)$, respectively, with minimum ionisation parameter $\log U \gtrsim -2$ (Sections 5.3 and 5.4). 

(2) The transient absorber densities are conservatively $\log n_{\rm H} ({\rm cm}^{-3}) \gtrsim 7$ based on the \hei*, \feii*, and \hi* lines, but much higher densities are likely present, with $\log n_{\rm H} ({\rm cm}^{-3}) \gtrsim 10$ indicated by the \nai~D lines.

(3) The density and ionization constraints place the transient absorber $\lesssim$0.4 pc from the quasar continuum source. However, we find that the \hi* absorption lines do not cover the Balmer broad emission-line region, suggesting that the actual location inside the BLR radius at $\lesssim$0.1 pc. At these smaller distances, the densities implied by the \feii* lines increase to $\log n_{\rm H} ({\rm cm}^{-3}) \gtrsim 8.7$ while the densities needed for \nai~D remain the same (Section 5.5). 

(4) Most of the transient lines are optically thick in spite of their weak appearance in the spectrum. Strong resonance lines of abundant ions have predicted optical depths $>$1000 to $\gtrsim$10$^5$ (Section 5.4.2 and Table 3). The weak \textit{observed} line strengths are due to partial covering of the quasar continuum source with covering fractions from $\sim$0.45 in strong UV resonance lines to $\sim$0.04 in \nai~D. This implies absorbing-region sizes $\lesssim$0.003 pc across (Section 5.6). 

(5) The range of covering fractions in different optically-thick lines points to absorption in a clumpy inhomogeneous outflow where the largest column densities and, perhaps, the highest densities occur in the smallest spatial volumes  (Section 5.6). 

(6) The radial thickness of the transient absorber indicated by the ratio $N_{\rm H}/n_{\rm H}$ is $\gtrsim$150 times smaller than the transverse sizes inferred from the covering fractions. This suggests that the transient absorber is flattened like a pancake in the radial direction or spatially distributed over a larger volume in many small clouds with individual sizes $<$$3\times 10^{13}$ cm (Section 5.6).

(7) The emergence of the optically thick transient lines is not compatible with simple ionisation changes in the outflow gas caused by changes in the quasar's ionising flux. We argue that the transient lines appeared due to transverse motions in the outflow whereby small, dense, high-column density clumps entered our lines of sight to the emission source (Section 5.7). 

(8) The outflow speed of the transient absorber |v|$\,\sim 1900$ \kms\ at its derived radial distance, conservatively $\lesssim$0.4 pc,  indicates that the gas is still gravitationally bound to the central black hole (Section 5.7).

\subsection{A Clumpy Inhomogeneous Absorber}

The transient absorber in PG~1411+442 is not simply an isolated outflow clump that wandered into our lines of sight. It is almost certainly physically related to the more persistent mini-BAL outflow based on 1) their similar velocity shifts, 2) the range of mini-BAL widths approaching the narrow transient lines in the lower ions, and 3) related variabilities, e.g., the mini-BALs became broader and somewhat deeper in 2015 when the transient system first appeared and then narrower again in subsequent observations when the transient lines weakened (Section 4 and HTRV2). The transient absorber appears to be a compact structure with the most extreme physical conditions in an overall clumpy inhomogeneous outflow that includes a wide range of physical conditions. 

Cartoon illustrations of clumpy inhomogeneous absorbers can be found in \cite{Hamann01} and \cite{Hamann04}. They depict higher column densities and perhaps higher densities occurring in smaller structures \citep[see also][]{deKool02, Arav05, Hamann11, Moravec17}. These clumpy inhomogeneous outflow structures provide a natural explanation for observations of ionisation-dependent and optical depth-dependent covering fractions. The reality of such structures is also supported by recent theoretical outflow models \citep[e.g.,][]{Giustini12, Matthews16, Waters17}. 

The outflow lines in PG~1411+442 provide numerous indications inhomogeneous partial covering. Highly saturated resonance lines in transient system like  \civ , \siiv , and \nv\ (Section 5.4.1, Table 3) are not clearly evident in the observed spectrum because they are overwhelmed by the broader and deeper mini-BALs. This is \textit{not} due to larger column densities in the mini-BAL gas. It is caused, instead, by larger covering fractions in the mini-BAL outflow (that produce deeper line troughs) across wider ranges in velocity (that produce broader line profiles). Photoionization models show that the \civ\ and low-abundance \pv\ mini-BALs form roughly together in fully-ionised gas (Figure 8); however, the \civ\ trough is broader and deeper than \pv\ because it is several hundred times more optically thick \citep[Figures 2 and 3, HTRV2,][and Herbst et al., in prep.]{Hamann98, Leighly11, Borguet13, Capellupo17, Moravec17}. The \hei* \lam 5876 line in the transient system also forms in ionized gas, but it is weaker and narrower than the \civ\ and \pv\ mini-BALs because this line requires more extreme conditions (higher densities and larger column densities, Figure 7) that occur only in the transient absorber in spatially smaller regions. Within the transient system, the lines requiring the most extreme conditions such as \nai~D and the upper-state \feii* lines have weaker absorption and smaller covering fractions than much more optically thick lines like \feii\ UV8 \lam 1608. 

\subsection{Column Densities, Distances \& FeLoBALs}

The PG~1411+442 outflow composed of the transient absorber and the mini-BALs gas together in 2015 resembles low-ionisation outflows observed in other AGN, such as the Seyfert 1 galaxy NGC 4151 \citep{Kraemer01,Crenshaw00} and FeLoBAL quasars \citep[e.g.,][]{deKool01, deKool02b, deKool02c, Hall02, Hall03, Hall07, Trump06, Ji15, Shi16b, Sun17}. The prototype FeLoBAL quasar, LBQS~0059$-$2735, has a similar combination of broad BALs in strong lines of higher ions such as \civ\ plus many narrow low-ionisation lines at low velocities including \feii* at energies reaching $\gtrsim$3 eV \citep[][Hamann et al., in prep.]{Hazard87, Wampler95}. These features in LBQS~0059$-$2735 also require densities $\log n_{\rm H} ({\rm cm}^{-3}) \gtrsim 7$ and column densities $\log N_{\rm H} ({\rm cm}^{-2}) \gtrsim 23$ within a few pc of this high-luminosity quasar \citep[see also][]{Shi16, Shi16b}. There are also FeLoBAL outflows with lower densities that, combined with lower derived column densities, appear to indicate much larger distances from the quasars \citep[e.g.,][]{deKool01, deKool02b, Moe09, Dunn10}. This has inspired theoretical models of FeLoBALs forming in situ at kpc distances, as fast quasar outflows shred and disperse interstellar clouds in the host galaxies \citep{Faucher12b}.  

FeLoBALs might indeed form in a range of locations with different physical origins in different quasars, but at least some of the range in reported locations is due to limitations in the data or uncertainties in the analysis. BAL outflows are often ranked by ionisation: HiBALs have only high-ionisation lines like \civ , LoBALs have those lines plus \mgii\ absorption, while FeLoBALs additionally include \feii\ and \feii* lines \citep[e.g.,][]{Hall02, Trump06, Gibson09}. Our finding in PG~1411+442 that the \feii\ and \feii* lines require an intense radiation field behind a thick layer of fully-ionised gas suggests that the main parameter driving the progression from HiBALs to LoBALs and FeLoBALs is larger total column densities, \textit{not} smaller ionisation parameters and larger distances from the quasars. This is supported by recent work indicating that BAL outflows typically have large ionisation parameters, $\log U \gtrsim -0.5$, and large column densities in ionised gas, $\log N_{\rm H} ({\rm cm}^{-2}) \gtrsim 22.7$, based on measurements of \pv\ BALs in median composite BAL quasar spectra (\citealt{Hamann19}, see also \citealt{Hamann99, Hamann02, Leighly11, Borguet13, Chamberlain15, Capellupo17, Moravec17}). The \cite{Hamann19} study shows further that the stronger low-ionisation lines in LoBALs are accompanied by stronger \pv\ absorption compared to HiBALs. This directly links LoBALs to \textit{high} ionisation parameters and large total column densities where low-ionisation absorption lines like \mgii\ form in radiatively-shielded regions behind large columns of fully-ionised \pv-absorbing gas \citep[see also][]{Baskin14}. 

These results indicate that some previous estimates of small total column densities in (Fe)LoBAL outflows (that then require small ionisation parameters and large distances from the quasars) need to be revised upward by considering the insidious nature of optical depth-dependent covering fractions in inhomogeneous outflows. In particular, studies that assume complete covering or adopt a single covering fraction for all lines can grossly underestimate the true line optical depths and column densities. For example, a single covering fraction used for the transient absorber in PG~1411+442 would need to be roughly $C_{\textrm{v}}\gtrsim 0.5$ based on the measured depths of the strongest UV lines (Table 2). Then, using that result, we would infer optical depths in weak lines that are only $\lesssim$2 times larger than their apparent values listed in Table 2, e.g., $\tau_o \lesssim 0.03$ for \feii* 74 \lam 6248 and $\lesssim$0.15 for \feii* 42 \lam 5169. However, the measured strengths and ratios of these lines indicate that at least \feii* 42 \lam 5169 is saturated. Their true covering fractions are $\lesssim$0.08 and their true optical depths are $\tau_o(\lambda 6248) \gtrsim 0.2$ and $\tau_o(\lambda 5169) \gtrsim 7$ (Section 5.4.2). Therefore, adopting $C_{\textrm{v}}\gtrsim 0.5$ would underestimate the line optical depths and total $N_H$ column densities by a factor of $\gtrsim$7 based on the measured \feii* 74 \lam 6248 line strength, $\gtrsim$47 based on \feii* 42 \lam 5169, or more if (like most FeLoBAL studies) only stronger \feii* lines from lower energy states are measured. 

The main takeaway (see also Section 6.1) is that inhomogeneous absorbers can produce optical depth-dependent covering fractions and optical depth-dependent line strengths in fully saturated lines. in inhomogeneous outflows require us to examine the weakest lines available (based on small $gf$ values, highly-excited states, or low abundances) to derive the most stringent and reliable constraints on the total column densities. The larger column densities inferred from this analysis then, in turn, indicate smaller distances from the quasars because they produce self-shielding in the outflow gas. 

\subsection{Is the Transient Absorber a BLR Outflow?}

Another interesting result is that the high densities and large column densities in the transient absorber are similar to quasar broad emission-line regions. Typical BLR densities might be closer to $\log n_{\rm H} ({\rm cm}^{-3}) \sim11$ \citep{Ferland92}, but there should also be a wide range \citep{Baldwin95, Korista97} given that the broad \ciii ] \lam 1909 emission line requires $\log n_{\rm H} ({\rm cm}^{-3}) \lesssim 9.5$ \citep{Davidson79}. The transient lines might identify a piece of the BLR that wandered into our line of sight. In the conventional picture of BLRs concentrated near the accretion-disk plane, the transient absorber could be an upper extension of the BLR farther above the disk, perhaps involved a BLR outflow \citep[as observed regularly via blueshifts in the \civ\ broad emission lines,][]{Leighly04b, Leighly07b, Richards11, Kruczek11, Veilleux16, Coatman16, Sun18}. The inferred location of the transient absorber $\lesssim$0.4 pc from the central light source is consistent with the predicted BLR radius of $\sim$0.1 pc in PG~1411+442 (Section 5.5). 

It is also interesting that the transient absorber moving at v$\;\sim -1900$ \kms\ is still gravitationally bound to the black hole (Section 5.7). A similar situation occurs in the FeLoBAL quasar LBQS~0059$-$2735, where narrow absorption lines at low speeds $\lesssim$500 \kms\ identify high-density high-column density clumps that also appear gravitationally bound \citep[][Hamann et al., in prep.]{Wampler95}. \cite{Shi16} report similar results for another FeLoBAL quasar. These gravitationally bound clumps might be the upper envelope of clumpy BLRs composed of material rising and falling above the accretion disk plane \citep{Czerny11, Czerny16, Baskin18}. 

In both PG~1411+442 and LBQS~0059$-$2735, the narrow low-velocity absorption-line clumps are accompanied by much broader and more blueshifted BALs or mini-BALs in higher ions like \civ. In LBQS~0059$-$2735, there is a clear progression from the narrow lines of \niii* and \feii*, which probe the highest densities and largest column densities at low velocities, to increasingly broader and more blueshifted absorption troughs in stronger transitions in higher ions, e.g., \mgii\ \lam 2796,2804, \aliii \lam 1855,1863, and \civ \lam 1548, 1551 (\citealt{Wampler95}, Hamann et al., in prep., see also \citealt{Hall02, Hall03} and \citealt{Shi16b, Shi16} for more examples). 

We might be witnessing in these quasars the development of high-speed outflows arising from the upper BLR, where dense clumps of BLR gas are accelerated and/or ablated and dispersed as they become exposed to stronger radiative forces higher above the accretion-disk plane \citep[see also][]{Wampler95b, Baskin13, Baskin15, Waters17}. These phenomena might be common in quasars but rare in observed quasar spectra because fortuitous timings and/or viewing perspectives are needed to measure them in absorption lines. We are currently monitoring both PG~1411+442 and LBQS~0059$-$2735 to search for absorption-line variabilities that might signal acceleration and/or clump dispersal related to this picture of the outflow origins (see HTRV2 and Hamann et al., in prep.). 

\section*{Acknowledgements}

We are grateful to the COS instrument team at the Space Telescope Science Institute (STScI) for their assistance in helping to confirm the reality of the peculiar ripples in our HST-COS spectrum of PG~1411+442 obtained in 2015. We specifically thank Justin Ely for guiding this effort at STScI and producing a detailed report on the COS instrument status and data handling. Revision 114 of the IFUDR GMOS data reduction package, based on the Gemini IRAF package, was provided by James Turner and Bryan Miller via the Gemini Data Reduction User Forum. This work was supported by grants from STScI in the guest observer programs HST-GO-12569, HST-GO-13451, HST-GO-13460, HST-GO-14460, and HST-GO-14885. S.V. acknowledges support from a Raymond and Beverley Sackler Distinguished Visitor Fellowship and thanks the host institute, the Institute of Astronomy, where this work was concluded. S.V. also acknowledges support by the Science and Technology Facilities Council (STFC) and by the Kavli Institute for Cosmology, Cambridge.

%%%%%%%%%%%%%%%%%%%%%%%%%%%%%%%%%%%%%%%%%%%%%%%%%%



\bibliographystyle{mnras}
\bibliography{../../bibliography}


% Don't change these lines
\bsp	% typesetting comment
\label{lastpage}
\end{document}